\documentclass{article} 
\usepackage{iclr2024_conference,times}

\usepackage{amsmath,amsfonts,bm}

\def\eqref#1{equation~\ref{#1}}

\def\1{\bm{1}}

\DeclareMathAlphabet{\mathsfit}{\encodingdefault}{\sfdefault}{m}{sl}
\SetMathAlphabet{\mathsfit}{bold}{\encodingdefault}{\sfdefault}{bx}{n}

\usepackage{hyperref}
\usepackage{amsmath}
\usepackage{url}
\usepackage{listings, booktabs}
\usepackage{graphicx}

\usepackage{wrapfig}
\usepackage[linesnumbered,ruled,vlined]{algorithm2e}
\usepackage{algorithmic}
\usepackage{multicol, multirow}
\usepackage{xcolor}
\usepackage{tabularx}
\usepackage{booktabs}
\usepackage[labelfont=bf,format=plain,justification=raggedright]{caption}
\usepackage{arydshln}

\SetKwBlock{DoParallel}{DoParallel}{end}

\SetCommentSty{mycommfont}
\usepackage{amsthm}

\theoremstyle{definition}

\definecolor{codegreen}{rgb}{0,0.6,0}
\definecolor{codegray}{rgb}{0.5,0.5,0.5}
\definecolor{codepurple}{rgb}{0.58,0,0.82}
\definecolor{backcolour}{rgb}{0.95,0.95,0.92}

\lstdefinestyle{mystyle}{
  backgroundcolor=\color{backcolour}, commentstyle=\color{codegreen},
  keywordstyle=\color{magenta},
  numberstyle=\tiny\color{codegray},
  stringstyle=\color{codepurple},
  basicstyle=\ttfamily\footnotesize,
  breakatwhitespace=false,         
  breaklines=true,                 
  captionpos=b,                    
  keepspaces=true,                 
  numbers=left,                    
  numbersep=5pt,                  
  showspaces=false,                
  showstringspaces=false,
  showtabs=false,                  
  tabsize=2
}

\usepackage[strict]{changepage}
\usepackage{xcolor}
\usepackage{framed}
\definecolor{demonstrationshade}{rgb}{0.95,0.95,1}
\definecolor{promptshade}{rgb}{0.95,0.95,1}

\usepackage{authblk}

\title{Interactive Speculative Planning: \\Enhance Agent Efficiency through Co-design of System and User Interface}

\author[*,1]{{Wenyue Hua}\thanks{Work during Microsoft Research internship. I'm very grateful for extensive discussion with Dujian Ding from University of British Columbia, Devang Acharya from Carnegie Mellon University in Qatar, Adam Fourney, Jaime Teevan, Brent Hecht, Pei Zhou, and other members of Microsoft Office of Applied Research.
}}
\author[2]{{Mengting Wan}}
\author[2]{{Shashank Vadrevu}}
\author[2]{{Ryan Nadel}}
\author[1]{{Yongfeng Zhang}}
\author[*, 3]{{Chi Wang}\thanks{Work done while working at Microsoft Research.}}
\affil[1]{Rutgers University, New Brunswick}
\affil[2]{Microsoft}
\affil[3]{Google Deepmind}
\affil[1]{\texttt{\{wenyue.hua, yongfeng.zhang\}@rutgers.edu}}
\affil[2]{\texttt {\{mengting.wan, svadrevu, Ryan.Nadel\}@microsoft.com}}
\affil[3]{\texttt {chi@chiwang.cc}}

\begin{document}

\maketitle

\begin{abstract}
Agents, as user-centric tools, are increasingly deployed for human task delegation, assisting with a broad spectrum of requests by generating thoughts, engaging with user proxies, and producing action plans. However, agents based on large language models (LLMs) often face substantial planning latency due to two primary factors: the efficiency limitations of the underlying LLMs due to their large size and high demand, and the structural complexity of the agents due to the extensive generation of intermediate thoughts to produce the final output. Given that inefficiency in service provision can undermine the value of automation for users, this paper presents a human-centered efficient agent planning method -- Interactive Speculative Planning -- aiming at enhancing the efficiency of agent planning through both system design and human-AI interaction. Our approach advocates for the co-design of the agent system and user interface, underscoring the importance of an agent system that can fluidly manage user interactions and interruptions. By integrating human interruptions as a fundamental component of the system, we not only make it more user-centric but also expedite the entire process by leveraging human-in-the-loop interactions to provide accurate intermediate steps. Code and data will be released. 
\end{abstract}

\section{Introduction}
Large language models (LLMs) have demonstrated strong reasoning abilities \citep{zhang2024llm, qiao2022reasoning, fan2023nphardeval, fan2024nphardeval4v, jin2024impact}, enabling them to plan and interact with external tools and the real world. This has led to the development of LLM-based agents, which have become popular as task solvers and human assistants. Various agent frameworks have been created to facilitate these applications, including single-agent systems such as Langchain \citep{topsakal2023creating}, OpenAGI \citep{ge2024openagi}, and HuggingGPT \citep{shen2024hugginggpt}, as well as multi-agent systems like AutoGen \citep{wu2023autogen}, MetaGPT \citep{hong2023metagpt}, BabyAGI \citep{nakajima2023babyagi}, and Camel \citep{li2023camel}. Numerous methods have also been proposed to enhance the performance of LLM-based agents, ranging from chain-of-thought \citep{wei2022chain}, tree-of-thought \citep{yao2024tree}, ReAct \citep{yao2022react}, Reflexion \citep{shinn2024reflexion}, to multi-agent discussion \citep{chan2023chateval} systems.

However, these high-performing advancements in agents often come at the expense of time efficiency \citep{zhou2024survey, ding2024occam, zhang2024you}, which can be attributed to three main reasons: (1) the underlying backbone language model can be inefficient due to its increasingly large size and high request volume, (2) the complex agent structure, such as tree-of-thought and ReAct, which requires generating prolonged thought before the final answer, leading to extended waiting time and increased token generation costs, and (3) the sequential nature of action steps in plans, where one action must be completed before the next can begin. But notice that not all steps in agent planning necessitate computationally intensive thought processes, making the universal application of complex agent architectures or agents with advanced backbone LLMs inefficient.

Moreover, latency is a critical factor for user experience. Many studies \citep{horvitz1999principles, barron2004graphical, simpson2007impact, carr1992effects} have demonstrated the physiological and psychological impacts from interaction delays during human-computer interaction on users. While agents are designed to assist users, few designers have prioritized user experience, which should be of high importance. In addition, numerous studies have discussed the role of automation in human-computer interaction \citep{lubars2019ask, hemmer2023human}, highlighting a low preference for full AI control in task delegation and a strong preference for machine-in-the-loop or human-in-the-loop designs where humans maintain a central role. Thus, a fully automated agent system with long intermediate delays is suboptimal for user experience, a feature that is, however, prevalent in most LLM-based agent systems today.


\begin{wrapfigure}{L}{0.5\textwidth}
\centering
\includegraphics[width=0.45\textwidth]{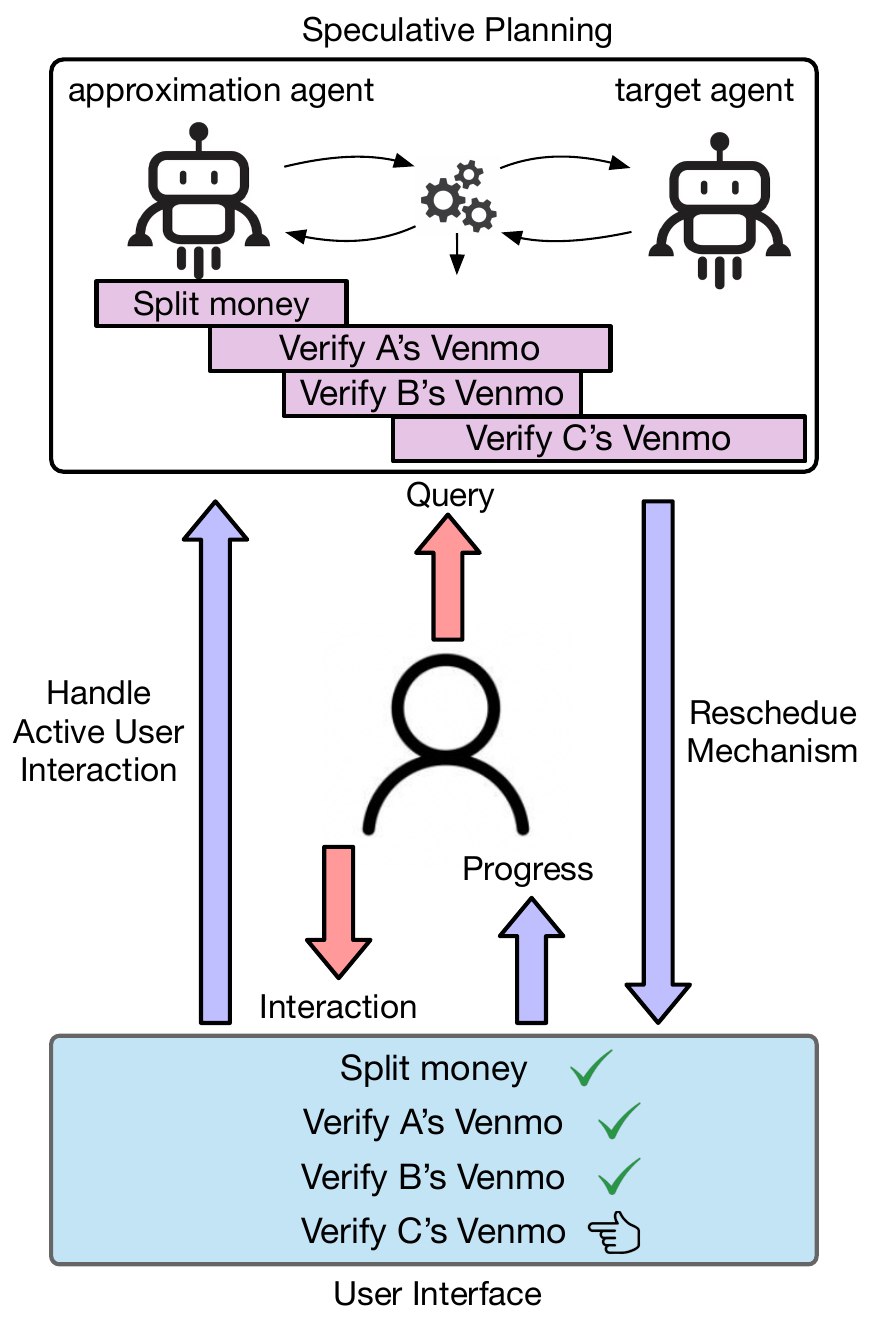}
\caption{Interactive Speculative Planning: user query is handled by speculative planning with approximation and target agent. Then a rescheduling mechanism serializes the computed result on UI and enables the user to actively interact with the system for further acceleration. Finger-pointed action is the action intervened by the user.}
\label{fig:demo}
\end{wrapfigure}

Therefore, this work aims to address the latency issue from both aspects of system design and human interaction by introducing an interactive efficient planning algorithm, representing the first system for agent latency efficiency and management of human interactions and interruptions: \textbf{Interactive Speculative Planning}. Figure \ref{fig:demo} is a simple domonstration of the system. This approach seamlessly integrates temporal efficiency and human-in-the-loop interaction, anticipating user engagement during periods of long latency. By treating user input as intermediate results, the system accelerates the overall process, thereby enhancing both temporal efficiency and user experience. Consequently, this system offers a more user-centric and efficient solution for agents as human delegates.

\begin{figure}
    \centering
    \includegraphics[scale=0.55]{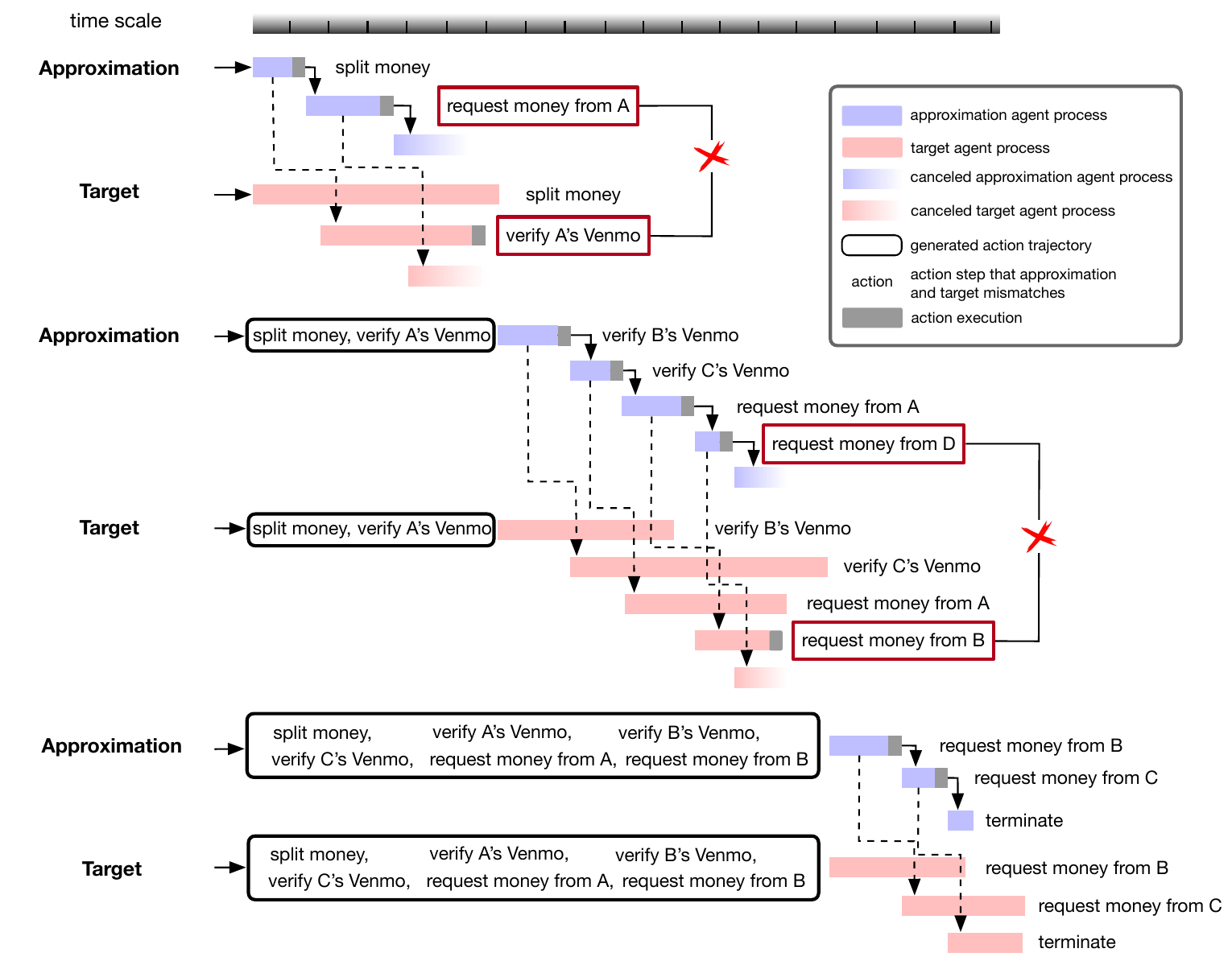}
    \caption{Speculative Planning Algorithm Demonstration, where the cross symbol indicates the step where the $\mathcal{A}$'s computed result differs from that of $\mathcal{T}$.}
    \label{fig:spec_plan}
\end{figure}

The system-level algorithm is speculative planning, which is inspired by speculative decoding \citep{leviathan2023fast, liu2023online, chen2023accelerating, spector2023accelerating, liu2024kangaroo, cai2024medusa}. It leverages two agent systems: an efficient but less capable approximation agent, and a slower but more powerful target agent. For each task, the approximation agent generates action steps sequentially. Simultaneously, for every step the approximation agent produces, the target agent is asynchronously called to generate the next step, using the current trajectory from the approximation agent as a provisional prefix. In this process, the calls to the approximation agent are sequential, while those to the target agent are asynchronous. For each action step, if the outputs of the approximation agent and the target agent match, the process continues. However, if there is a mismatch, the approximation agent is halted, and its output is replaced by the target agent's output to ensure performance is not compromised. Figure \ref{fig:spec_plan} presents the process.

This strategy potentially reduces the time a target agent takes for to complete the task to that of the approximation agent, thereby enhancing time efficiency. It should be noted that while we consider a single user interface of the agent system, the system backend can be built with various architectures, including a multi-agent design.

\begin{figure}[!ht]
    \centering
    \includegraphics[scale=0.7]{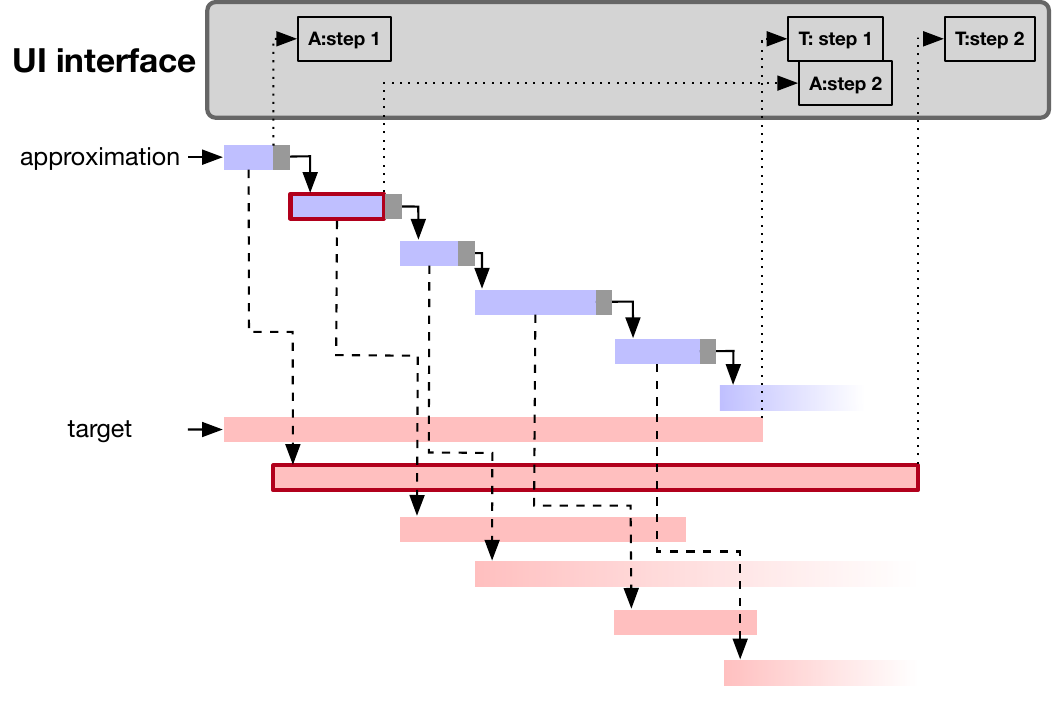}
    \caption{User interface which guarantees a sequential presentation of approximation agent's output and target agent's output with minimum perceived latency.}
    \label{fig:reschedule}
\end{figure}

Note that under the speculative planning algorithm, target agent calls are asynchronous, leading to non-sequential outputs. To facilitate user interaction, we design a UI-level rescheduling algorithm \citep{oh2024exegpt, cheng2024slice, mei2024llm, jawahar2023llm, srivatsa2024preble} that presents both the approximation agent's results and the target agent's results sequentially and clearly, as illustrated in Figure \ref{fig:reschedule}. The sequential presentation of the outputs enables users to accurately perceive the computation latency imposed by the target agent. Consequently, users may intervene in the process at their discretion, such as when a computation step is prolonged or yields erroneous results. This approach differs from traditional speculative decoding, where an algorithm judges whether to accept the results from an approximation agent, often based on probability distributions. In contrast, Interactive Speculative Planning enables active human engagement in the decision-making process, allowing them to interrupt lengthy processes and evaluate whether to accept or complement the algorithm's results. This human-in-the-loop approach makes the system user-centric and efficient.

In summary, with active user intervention, Interactive Speculative Planning can be viewed as an interactive framework involving \textbf{three} agents: the approximation agent, the target agent and the human agent. These three agents collaborate and interleave their operations to collectively accelerate the overall agent planning process.

In the rest of the paper, Section \ref{sec:related} will introduce related work in agent and agent efficiency, Section \ref{sec:ISP} introduces the algorithm, Section \ref{sec:analysis} provides theoretical analysis on the time, rate limit, and total token generated, as well as simulation experiment, Section \ref{sec:experiment} provides empirical experiment result on actual datasets, Section \ref{sec:limitation} discusses the current limitations and potential future work of the algorithm, and Section \ref{sec:conclusion} concludes the paper.

\section{Related Work}
\label{sec:related}
Various agent systems \citep{xi2023rise, liu2023dynamic, ge2023llm} have been developed, including single agent such as Hugginggpt \citep{shen2024hugginggpt}, OpenAGI \citep{ge2024openagi}, and BabyAGI \citep{nakajima2023babyagi}, and multi-agent systems \citep{du2023improving} such as AutoGen \citep{wu2023autogen, zhang2024training} and Camel \citep{li2023camel}, based on the strong reasoning ability \citep{wu2024mathchat, zhang2023ideal} and common sense knowledge \citep{kwon2024toward} encoded in LLMs. To improve the performance of LLM-based agents, various methods have been proposed. The most basic approach is the chain-of-thought \citep{wei2022chain}, where the LLM generates a step-by-step thought process for each action. More advanced methods include ReAct \citep{yao2022react}, where the agent thinks before acting, and Reflexion \citep{shinn2024reflexion}, where the agent thinks, acts, and reflects on its decisions. The tree-of-thoughts \citep{yao2024tree} method involves the agent thinking several steps ahead before acting. In addition, multi-agent discussion systems \citep{du2023improving, hua2023war, lin2024battleagent, wu2023autogen} have been developed in which multiple agents discuss and debate to improve performance. In general, it is observed that stronger backbone models and more complex multi-LLM interaction usually lead to better agents \citep{wang2024rethinking, li2024more, chen2024more}. 

However, these improvements in agent performance often come at the expense of time efficiency, as longer thought processes result in extended waiting times. Although the agent task can be intricate and sometimes only the most powerful models may be capable of executing them effectively as suggested by \citep{xie2024travelplanner}, not all steps within a task are equally challenging to plan and generate \citep{zhang2023ecoassistant, saha2024system}. Therefore, a dynamic selection of appropriate LLMs for specific tasks can be a viable strategy to balance performance and efficiency/cost. 

Numerous methods have been developed to enhance either cost or time efficiency \citep{zhang2023ecoassistant, ding2024hybrid, saha2024system}. EcoAssistant \citep{zhang2023ecoassistant} is the first system aimed at cost-efficient agents, initiating tasks with the most economical agent and switching to more capable and expensive agents only upon failure of the cheaper alternative. The System-1.x Planner \citep{saha2024system} introduced a controllable planning framework using language models, capable of generating hybrid plans and balancing between complex and simple agent planning strategies based on problem difficulty, potentially offering both time and cost efficiency. However, the System-1.x Planner is limited to specific planning strategies and requires extensive training. In contrast, our proposed Interactive Speculative Planning can adopt any combination of approximation and target agent in a training-free manner, guaranteeing performance that is at least equivalent to, and potentially superior to (with user interventions) that of the target agent alone.

\section{Interactive Speculative Planning}
\label{sec:ISP}
Interactive Speculative Planning is a collaborative framework that enhances the efficiency and accuracy of agent planning by integrating the efforts of three agents: the approximation agent, the target agent and the human agent. The approximation agent generates quick but potentially inaccurate steps, while the target agent verifies and refines these steps. The human agent intervenes to correct or optimize the latency of the planning process, ensuring that the final plan meets user expectations. This interactive approach accelerates the overall planning process and improves user experience by reducing latency and allowing for real-time adjustments.

In this section, we delves into the algorithm: Section 3.1 introduces speculative planning and Section 3.2 introduces what user interactions are expected and how the system incorporates user interactions. Let us denote the approximation agent by $\mathcal{A}$ and the target agent by $\mathcal{T}$.

\subsection{Speculative Planning}
\label{subsec:sp}

\begin{wrapfigure}{L}{0.5\textwidth}
\centering
\includegraphics[width=0.5\textwidth]{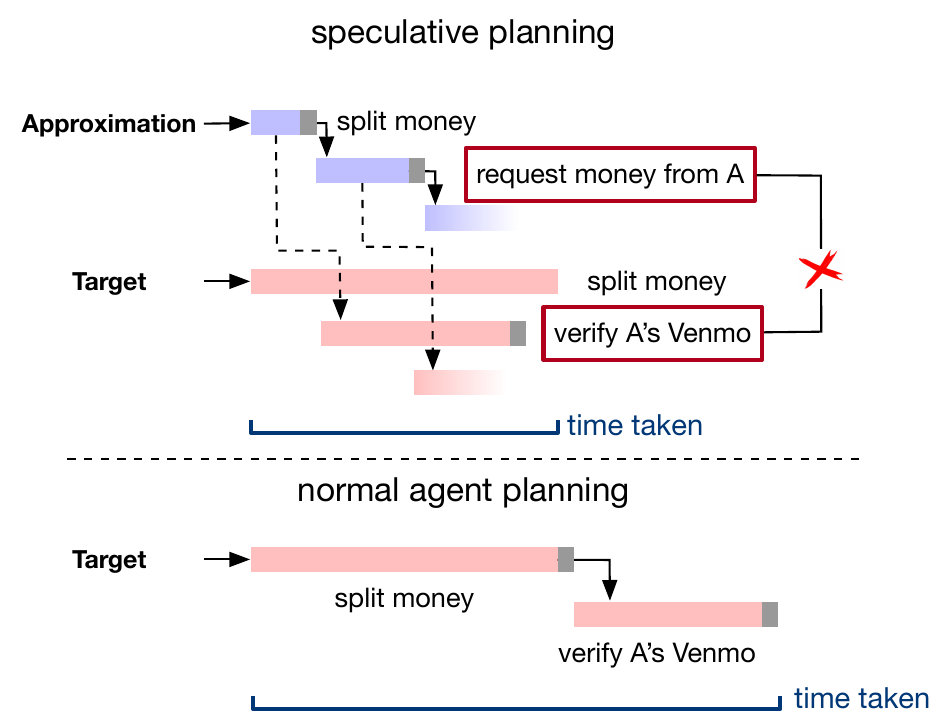}
\caption{Comparing the time taken to generate the first two steps of a task by agent system using speculative planning and normal agent planning.}
\label{fig:comparison}
\end{wrapfigure}
The core concept of how time is saved by speculative planning is to expedite agent planning by employing a fast and efficient approximation agent $\mathcal{A}$ to resolve the task sequentially. For every length-$i$ prefix of the step generated by $\mathcal{A}$, both $\mathcal{A}$ and $\mathcal{T}$ are run simultaneously to generate the $i+1$th step based on $\mathcal{A}$'s action history, without waiting for $\mathcal{T}$ to finish the 
$i$th step. If the $i$th step of the plan generated by both agents matches after $\mathcal{T}$ finishes it, then the more efficient but less capable agent $\mathcal{A}$ is deemed to have correctly computed the step, and $\mathcal{T}$'s $i+1$th step computed based on it is usable. Time is thus saved because the time for $\mathcal{T}$ to compute steps $i$ and $i+1$ is reduced to the time taken by $\mathcal{A}$ to compute step $i$ and $\mathcal{T}$ to compute step $i+1$. However, if there is a mismatch, it implies that $\mathcal{A}$ has erred at the $i$th step, and its output is replaced with $\mathcal{T}$'s result. Furthermore, all concurrent calls of $\mathcal{A}$ and $\mathcal{T}$ with prefixes longer than $i$ must be halted and discarded, as they are based on an incorrect prefix and their results are unusable. In short, this algorithm achieves time savings by having $\mathcal{T}$ utilize the result generated by the fast $\mathcal{A}$ as a prefix to generate the next step, rather than waiting for prefix steps from the slower $\mathcal{T}$ to be completed.

This algorithm achieves time savings by having $\mathcal{T}$ utilizes the result generated by the fast $\mathcal{A}$ as a prefix to generate the next step, rather than waiting for prefix steps from the slower $\mathcal{T}$ to be completed. An example comparison of the time taken to generate the first two steps of a task using speculative planning and normal agent planning is illustrated in Figure \ref{fig:comparison}: 

Figure \ref{fig:spec_plan} presents a scenario where one person with their friends A, B, C went to a restaurant and they paid the bill, and now they need to split the money with their friends. When initiating speculative planning, both $\mathcal{A}$ and $\mathcal{T}$ are started simultaneously to generate the first step. Upon $\mathcal{A}$'s completion of the first step (``split money''), both agents are called again simultaneously, utilizing ``split money'' as the current action trajectory to generate next step. This process is repeated for subsequent steps. Once $\mathcal{T}$ completes its first call, generating the first step (``split money''), the correctness of $\mathcal{A}$'s first step can be confirmed. Since the first step is correct, the second step from $\mathcal{A}$ is potentially correct waiting to be confirmed by the second step from $\mathcal{T}$, which is definitively usable. However, if $\mathcal{A}$'s output mismatches with $\mathcal{T}$'s output, all subsequent steps are deemed incorrect. In this example, $\mathcal{T}$ completes its second call (``verify A's Venmo'') before the first call which verifies the second step from $\mathcal{A}$ to be incorrect (``request money from A''). Consequently, all subsequent steps based on the action trajectory containing the second step are rendered useless. This includes the third call of $\mathcal{T}$ and the third step of $\mathcal{A}$ and so on.

To prevent an excessive number of concurrent target agent processes, the Interactive Speculative Planning algorithm introduces a hyperparameter $k$. This parameter sets a limit on the maximum number of steps that $\mathcal{A}$ that can sequentially propose and being executed before all corresponding target agent processes are completed. By controlling the value of $k$, users can flexibly manage the maximum number of concurrent target agent processes. A very simplified version of the speculative planning algorithm is presented in Algorithm \ref{algo:speculative_planning}:

\begin{algorithm}[!ht]
\DontPrintSemicolon 
\KwIn{Approximation agent: $\mathcal{A}$, Target agent: $\mathcal{T}$, task prompt $p$, action trajectory $\mathcal{S} = []$, \textsc{TEMINATE}=False, max approximation steps $k$}
\KwOut{$\mathcal{S}$}
i = 0\;
\While{not \textsc{TEMINATE}}{
  approximation$\_$step = 0\;
  \For{approximation$\_$step $\leq k$}{
      \DoParallel{
            create async process $\textsc{approximation}_i = \mathcal{A}(p, \mathcal{S})$ \; \tcp*{will return $i$-th action step $a_i$ by running $\mathcal{A}$}
            approximation$\_$step += 1\;
            create async process $\textsc{target}_i = \mathcal{T}(p, \mathcal{S})$ \; \tcp*{will return $i$-th action step $t_i$ by running $\mathcal{T}$}
        } 
      $a_i = $ await \textsc{approximation}$_i$ \; \tcp*{wait for $\mathcal{A}$ to finish computation sequentially}
      $o_i = $ \textsc{execution}($a_i$) \; \tcp*{execute the generated plan step $a_i$ and obtain observation $o_i$}
      update $p$ by adding description about $a_i$ and $o_i$\;
      i += 1 \;
      $\mathcal{S}$.append([$a_i$, $o_i$]) \; \tcp*{cache generated step $a_i$ and corresponding observation $o_i$}
      \For{$j = 0 \text{ to } i$}
        {
          \If{ $t_j \text{ is computed } \And t_j \neq a_j$} {
                $o_j' = $ \textsc{execution}($t_j$) \; \tcp*{re-execute the generated plan step $t_j$ and obtain observation $o_i'$}
                $\mathcal{S} = \mathcal{S}[:j] + [[t_j, o_j']]$ \; \tcp*{update cache}
                update $p$ by modifying $i$-th step based on description about $t_j$ and $o_i'$\;
                cancel all ongoing $\textsc{approximation}$ processes and $\textsc{target}_k$ if $k > j$ \; \tcp*{cancel useless processes}
                break the outer for loop and go to line 4\;
            }
        } 
      \If{ $\mathcal{S}[-1][0]$ is ``terminate''}{
        \textsc{TEMINATE}=True \;
        break the outer for loop\;
      }
   }
}
\Return $\mathcal{S}$
\caption{Speculative Planning Algorithm.}
\label{algo:speculative_planning}
\end{algorithm}

\subsection{UI Interaction Algorithm}

\begin{figure}[!ht]
    \centering
    \includegraphics[scale=0.65]{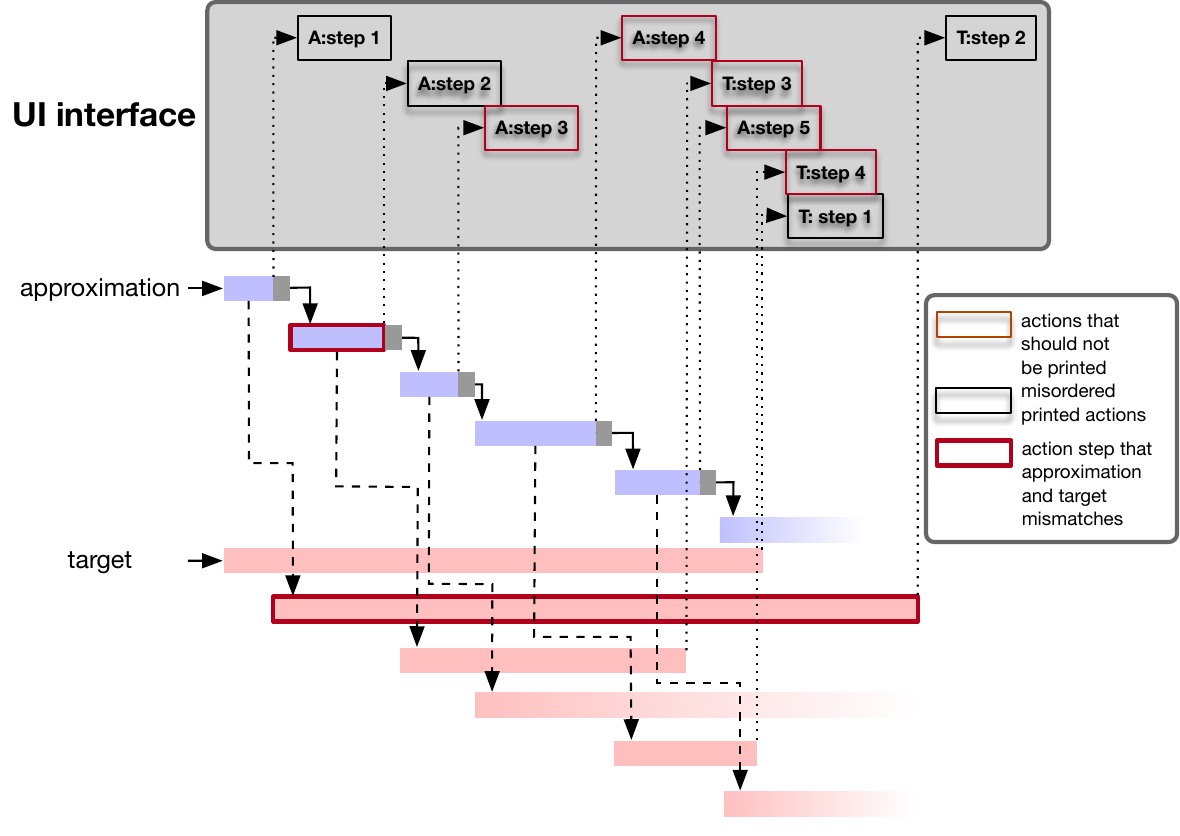}
    \caption{UI interface issues stemming from immediate presentation of computed action steps.}
    \label{fig:UI_issue}
\end{figure}

Now we present the interaction component of Interactive Speculative Planning. The user interface (UI) serves a two-fold goal: (1) from the aspect of perception, it aims to provide the user with an easy-to-follow result and a basic understanding of the algorithm's inner workings, allowing the user to see $\mathcal{T}$'s computation time for each step and how $\mathcal{A}$ is saving time; (2) from the aspect of interaction, it aims to provide system support for the user to actively interact with or interrupt the ongoing agent processes -- when $\mathcal{T}$ is taking too long for a step or neither $\mathcal{A}$ nor $\mathcal{T}$ provides a satisfying step proposal during generation. Therefore, the UI interface, together with the underlying system mechanism design, primarily addresses two key aspects: (1) what the users should see and (2) how the system can handle user interactions.

For the first goal, notice that immediately printing the outputs of $\mathcal{A}$ and $\mathcal{T}$ upon generation can be very confusing for two reasons: (1) some outputs of $\mathcal{A}$ and $\mathcal{T}$ should not be shown to the user at all, and (2) the outputs of $\mathcal{T}$ are misordered. Figure \ref{fig:UI_issue} presents an example scenario for the two issue. For issue 1: $\mathcal{A}$'s output on the second step of the plan mismatches with $\mathcal{T}$'s output, and thus all results generated by $\mathcal{A}$ based on the mistaken ``step 2'' will ultimately be discarded. However, an immediate output of the agent's generation will present $\mathcal{A}$'s computed steps ``step 3, 4, 5'' and $\mathcal{T}$'s computed steps ``step 3, 4'' which are generated based on the wrong prefix. For issue 2: as all $\mathcal{T}$'s calls are asynchronous, the time for each step to finish will not follow a sequential order, and thus an immediate printing out of the generated output will not be sequential either. Therefore, a rescheduling mechanism is needed to provide a clear presentation of the algorithm.

To ensure an understandable user interface to track the agents' progress and facilitate user intervention, the presented output is rescheduled by a Reschedule Mechanism. This mechanism allows the user to view verified and to-be-verified computed steps of $\mathcal{A}$ and $\mathcal{T}$ with minimal perceived latency. The Reschedule Mechanism, shown in Algorithm \ref{algo:UI}, takes the queue of $\mathcal{A}$ processes and the queue of $\mathcal{T}$ processes as input, tracing the last printed out message from either $\mathcal{A}$ and $\mathcal{T}$, and then decide which message to present next to the user: (1) it presents the $i$th step from $\mathcal{A}$ only after all preceding steps from $\mathcal{A}$ have been confirmed to be consistent with $\mathcal{T}$, ensuring that no steps computed based on unverified prefixes are presented (2) it presents the $i$th step from $\mathcal{T}$ only after all preceding steps from $\mathcal{T}$ have been presented, ensuring a sequential order. This design not only ensures a sequential presentation but also highlights the time difference between $\mathcal{A}$ and $\mathcal{T}$, allowing the user to identify which action is bottlenecking the program.


\begin{wrapfigure}{L}{0.7\textwidth}
\centering
\includegraphics[scale=0.55]{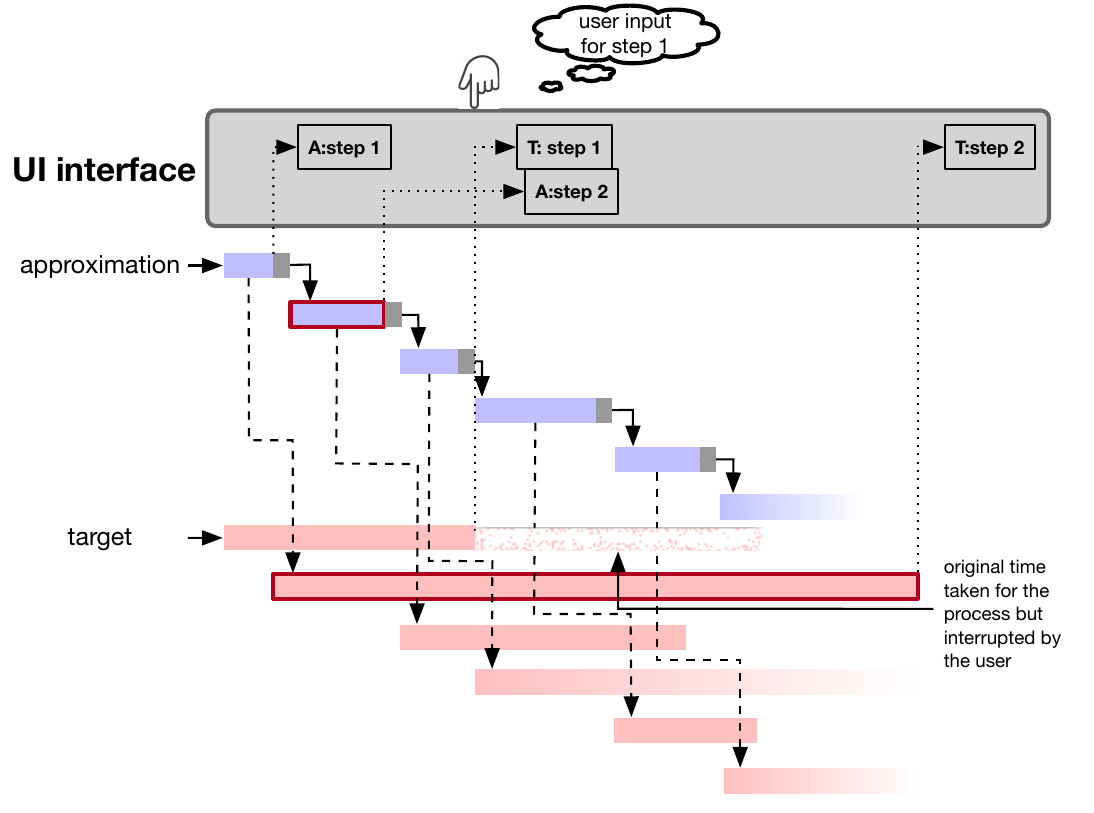}
\caption{How to handle user interruption during $\mathcal{T}$'s computation of step 1 due to excessive latency.}
\label{fig:user_interruption}
\end{wrapfigure}

For the second goal, we enable users to \textit{actively} interrupt the program at any time. Unlike current user interface designs in various agent systems \citep{wu2023autogen} where users are allowed to interact with the system when being \textit{passively} prompted to input information or opinions, we believe that users are more inclined to actively engage in the agent task delegation process \citep{lubars2019ask}. We handle user interaction in two common scenarios: (1) when noticing excessive perceived latency between the last presented output of $\mathcal{A}$ and the next output of $\mathcal{T}$ (assuming $\mathcal{A}$'s generation speed is sufficiently fast that users would not typically interrupt it), and (2) when dissatisfied with the outputs of $\mathcal{T}$ for a given step.

For the first scenario, since the UI interface presentation for the $i$-th step of the plan can indicate the latency $l_i$ between the presentation of the $i$-th approximation output $a_i$ computed by $i$-th process of $\mathcal{A}$ and the $i$-th target output $t_i$ computed by $i$-th process of $\mathcal{T}$, users can choose to interrupt during the time of $l_i$ and input their own value. The underlying system will handle this keyboard interruption by halting the $i$-th process of $\mathcal{T}$, incorporating the user's input into the agent action trajectory, while allowing all other concurrent processes to continue. Figure \ref{fig:user_interruption} demonstrates an example where the user interrupts the process after the presentation of $a_1$ due to excessive waiting time for $t_1$.

In the second scenario, users are able to interrupt the program if they deem the results from $\mathcal{T}$ unsatisfactory for a given step. During the brief presentation of the output $t_i$ for any step $i$, users can intervene and input their preferred optimal step for step $i$ as an oracle. Additional user interruption features, such as handling user suggestions instead of oracle results, or backtracking to previous steps rather than focusing on the current step, are potential avenues for future research.

\begin{algorithm}
\DontPrintSemicolon 
\SetAlgoLined
\SetKwProg{try}{try}{:}{}
\SetKwProg{catch}{catch}{:}{end}
\SetKwFunction{FMain}{Register-Handler}
\SetKwFunction{Fsub}{Exit-Handler}
\SetKwProg{Fn}{Function}{:}{}
\Fn{\FMain{target$\_$tasks, target$\_$task$\_$id}}{
    \Fn{\Fsub{signum, frame, target$\_$tasks, target$\_$task$\_$id}}{   
    target$\_$tasks[target$\_$task$\_$id].cancel()\;
    }
    \textbf{End Function}\;
    signal.signal(signal.SIGTSTP, partial(\Fsub, target$\_$tasks, target$\_$task$\_$id))\;
    $t_{target\_task\_id}$ = user input for action step target$\_$task$\_$id \;\tcp*{prompt user input to as oracle result}
    $ts$.append($t_{target\_task\_id}$)\;
}
\textbf{End Function}\;
\KwIn{Approximation process queue: $\mathcal{A}s$, Target process queue: $\mathcal{T}s$, Approximation result presentation index tracker $a\_tracker$, Target result presentation index tracker $t\_tracker$, Approximation result list $as$, Target result list $ts$}
\KwOut{$a\_tracker$, $t\_tracker$, $as$, $ts$}
\If{ $a\_tracker\leq t\_tracker$}{
    $i = t\_tracker$\;
    \If{ process $\mathcal{A}s[i]$ is completed}{
        present $as[i]$ to user interface\;
        Register-Handler(target$\_$tasks=$\mathcal{T}s$, target$\_$task$\_$id=$t\_tracker$)\; \tcp*{setup signal handler to enable proper handling of user interruption when user is waiting for $ts[i]$ to be computed}
        $a\_tracker$ += 1\;
    }
}
\Else{
    $i = a\_tracker$\;
    \If{ process $\mathcal{T}s[i]$ is completed}{
        present $ts[i]$ to user interface but allow user to modify $ts[i]$ as $ts[i]'$\; \tcp*{enable user to directly change $\mathcal{T}$'s computed result $ts[i]$ after presenting it to user}
        $t\_tracker$ += 1\;
    }
}
\Return $a\_tracker$, $t\_tracker$, $as$, $ts$
\caption{Rescheduling Mechanism with User Interruption.}
\label{algo:UI}
\end{algorithm}

\section{Efficiency Analysis}
\label{sec:analysis}
In this section, we will provide a theoretical analysis of the time savings (latency), total token generation requirement, and concurrent API call rate required by the speculative planning approach. Additionally, we will present simulated experiment results to support our analysis and demonstrate the effectiveness of the proposed method.

\begin{table}
\begin{tabularx}{\textwidth}{ll}
\toprule
  $n$ & the number of planning steps for a task  \\
  \cdashline{1-2}[.4pt/1pt]
  $time(\mathcal{A}, s)$ & the time the approximation agent $\mathcal{A}$ takes to generate step $s$ in the plan \\
  \cdashline{1-2}[.4pt/1pt]
  $time(\mathcal{T}, s)$ & the time the target agent $\mathcal{T}$ takes to generate step $s$ in the plan \\
  \cdashline{1-2}[.4pt/1pt]
  $e(s)$ & the time to execute a step $s$ in the plan and return an observation \\
  \cdashline{1-2}[.4pt/1pt]
  $token(\mathcal{A}, s)$ & the token the approximation agent $A$ requires to generate step $s$ in the plan \\
  \cdashline{1-2}[.4pt/1pt]
  $token(\mathcal{T}, s)$ & the token the target agent $T$ requires to generate step $s$ in the plan \\
  \cdashline{1-2}[.4pt/1pt]
  \multirow{2}{*}{$start\_time(\mathcal{A}, s_i)$} & it equals to $\Sigma_{j=b+1}^{j=i-1}(time(\mathcal{A}, s_j) + e(s_j))$, which indicates\\
  ~ & the start time of $\mathcal{A}$ to generate step $s_i$ since the one previous breaking point $b$ \\
  \cdashline{1-2}[.4pt/1pt]
  \multirow{3}{*}{$start\_time(\mathcal{T}, s_i)$} & it equals to $\Sigma_{j=b+1}^{j=i-1}(time(\mathcal{A}, s_j) + e(s_j))$ which indicates\\
  ~ & the start time of $\mathcal{T}$ to generate step $s_i$ since the one previous breaking point $b$. \\
  ~ & Notice that $start\_time(\mathcal{T}, s_i) = start\_time(\mathcal{A}, s_i)$\\
  \cdashline{1-2}[.4pt/1pt]
  \multirow{2}{*}{$end\_time(\mathcal{A}, s_i)$} & it equals to $\Sigma_{j=b+1}^{j=i}(time(\mathcal{A}, s_j) + e(s_j))$, which indicates\\
  ~ & the end time of $\mathcal{A}$ of generate step $s_i$ since the one previous breaking point $b$\\
  \cdashline{1-2}[.4pt/1pt]
  \multirow{2}{*}{$end\_time(\mathcal{T}, s_i)$} & it equals to $\Sigma_{j=b+1}^{j=i-1}(time(\mathcal{A}, s_j) + e(s_j)) + time(\mathcal{T}, s_i)$, which indicates\\
  ~ & the end time of $\mathcal{T}$ of generate step $s_i$ since the one previous breaking point $b$\\
\bottomrule
\end{tabularx}
\caption{Notation Summary}
\label{tab:summary}
\end{table}

\subsection{Latency Analysis}
This subsection analyzes the latency improvement brought by the speculative planning algorithm. We summarize the notations in Notation Summary \ref{tab:summary}.

When we do not utilize speculative planning, the time taken to generate and execute the whole plan is $\Sigma_{i\leq n} (time(\mathcal{T}, s_i) + e(s_i))$. To compute the time when employing speculative planning, we first define the list of breaking steps $B$, which consists of indices $i$ of steps $s$ in the plan where the sequential generation of $\mathcal{A}$ is halted, \emph{i.e.} when $\mathcal{A}$'s prediction $a_i=\mathcal{A}(i)$ differs from $\mathcal{T}$'s prediction $t_i=\mathcal{T}(i)$ for the $i$-th step in the planning, as well as when the number of continuous speculative steps generated by the approximation process reaches the hyperparameter $k$. \textit{For notational convenience, let's add $-1$ as the first element and $n-1$ as the last element in the list $B$}.

The time taken to generate and execute the entire plan is then determined by the following equation, where the time to compute each sequence of steps between two consecutive elements $B_i$ and $B_{i+1}$ in $B$, which is determined by step $i$ that takes longest time to compute for the target agent $\mathcal{T}$:
\begin{align}
    \Sigma_{B_i\in B[:-1]} (\max\{({end\_time(\mathcal{T}, s_j)}\mid B_{i}+1 \leq j \leq B_{i+1} \})
\end{align}

\begin{figure}[!ht]
    \centering
    \includegraphics[scale=0.75]{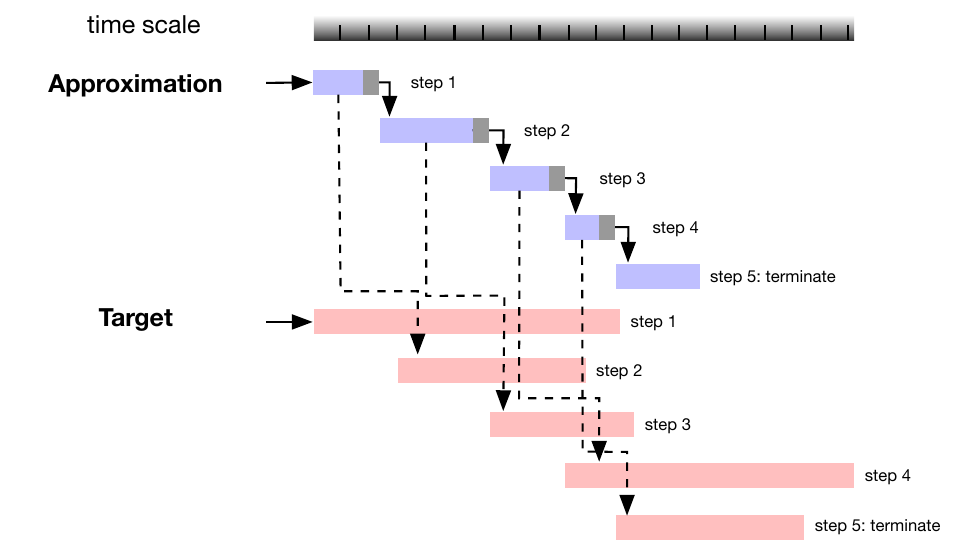}
    \caption{Best case scenario, assuming $k = n$}
    \label{fig:best_example}
\end{figure}

\textbf{Best case scenario} is that no step generated by $\mathcal{A}$ differs from the step generated by $\mathcal{T}$, as shown in Figure \ref{fig:best_example}. Thus in this specific case, the breaking points $B$ is simply all numbers $i$ smaller than $n$ such that $i\;\mathbf{mod}\;k$ = 0, and the computing time for the best case is:
\begin{align}
    \Sigma_{i\in \{i\;\mathbf{mod}\;k = 0\mid i < n\}} (\max_{i \leq j < i+k}end\_time(\mathcal{T}, s_j)))
\end{align}

\begin{figure}[!ht]
    \centering
    \includegraphics[scale=0.65]{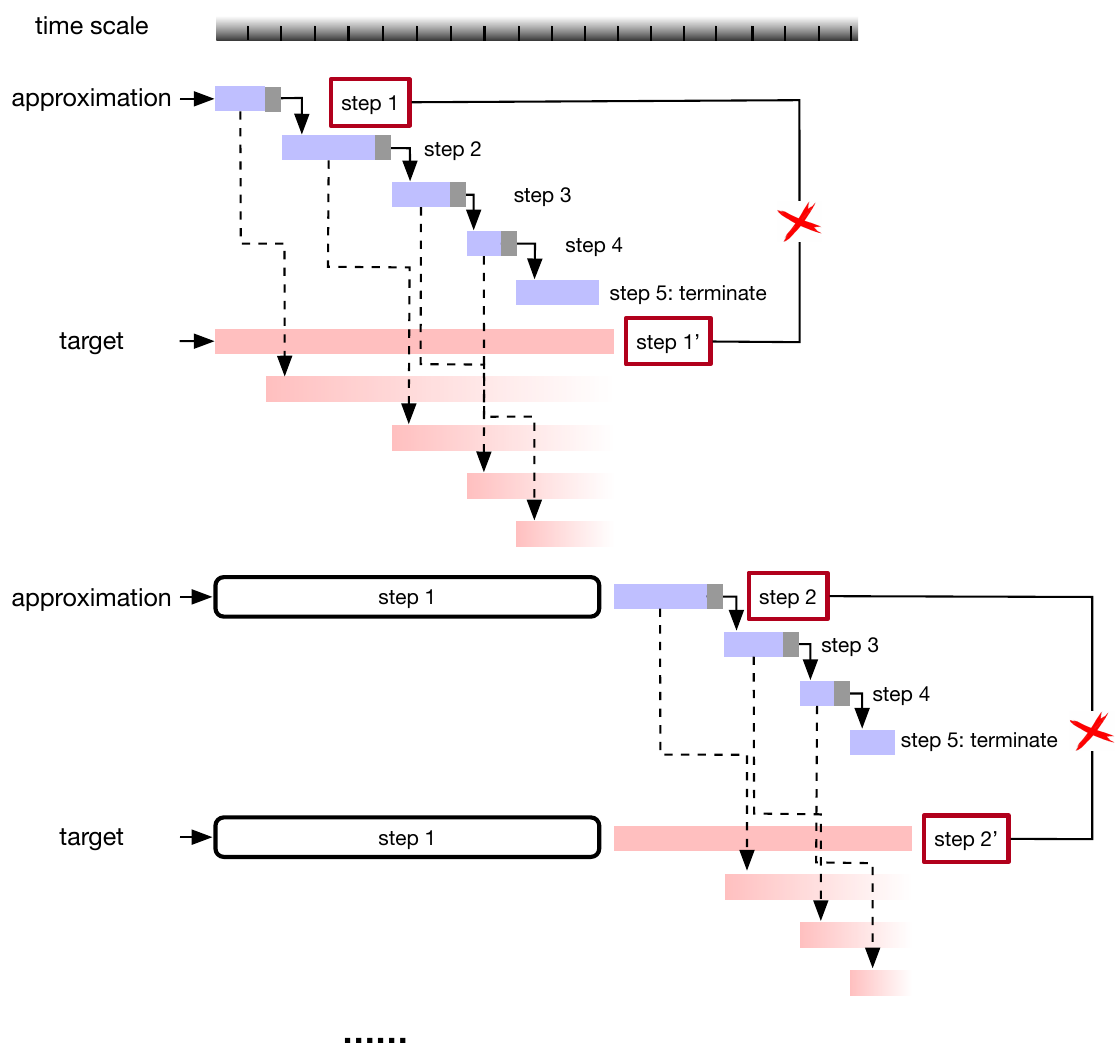}
    \caption{Worst case scenario}
    \label{fig:worst_example}
\end{figure}

\textbf{Worst case scenario} is that all steps generated by $\mathcal{A}$ are rejected by $\mathcal{A}$. A partial example is presented in Figure \ref{fig:worst_example}. In this extreme case, the set of breaking steps, $B$ comprises all integers from 0 to $n-1$.

Under these circumstances, the time taken to generate and execute the plan downgrades to normal agent planning. This equation calculates the sum of the time taken to generate and execute each step in the plan sequentially, without any speculative planning. The total time can be expressed as:
\begin{align}
\Sigma_{0\leq i\leq n-1} (time(\mathcal{T}, s_i) + e(s_i))
\end{align}

The aforementioned worst-case scenario demonstrates that, in terms of time efficiency, speculative planning is upper-bounded by the time taken in non-speculative planning. This implies that the maximum time required for speculative planning will not exceed the time taken by the traditional, non-speculative approach.

\subsection{Total token required} 
In this subsection, we analyze the total token generation when using the speculative planning algorithm. 

When not utilizing speculative planning, the total number of tokens used to generate and execute the plan is $\Sigma_{i< n} token(\mathcal{T}, s_i)$. Speculative planning requires more tokens, as both $\mathcal{A}$ and $\mathcal{T}$ go through the entire plan at least once, potentially generating ``wasted'' tokens -- proposed steps that are not used in the final plan which are computed based on incorrect prefix. Between any two breaking points $B_i$ and $B_{i+1}$, the number of tokens generated is the sum of tokens generated by $\mathcal{A}$ and $\mathcal{T}$ for steps $s_j$ in between as well as ``unused/wasted'' tokens generated by both $\mathcal{A}$ and $\mathcal{T}$ for any step $s_j$ such that $j \geq B_{i+1}$, where the process ends before $\mathcal{T}$ finishes all the process between any two breaking points $B_i$ and $B_{i+1}$. Thus, we represent the tokens generated $T_{B_i}$ between two consecutive breaking points $B_i$ and $B_{i+1}$ as below:
\begin{align}
    T_{B_i} &= \underbrace{\Sigma_{j = B_i+1}^{j = B_{i+1}} (token(\mathcal{A}, s_j) + token(\mathcal{T}, s_j))}_{\text{ sum of tokens generated by }\mathcal{A} \text{ and } \mathcal{T} \text{ in between } B_i \text{ and } B_{i+1}} + \underbrace{\Sigma_{j = B_{i+1}+1}^{M_i} (token(\mathcal{A}, s_j) + token(\mathcal{T}, s_j))}_{\text{wasted tokens}}
\end{align}
where $Q = \max_{B_{i} < l\leq B_{i+1}}\{end\_time(\mathcal{T}, s_l)\}$ is the ending time for all steps between $B_i$ and $B_{i+1}$ to be computed, and $M_i = \min\{\max\{l< n\mid end\_time(\mathcal{A}, s_l) \leq Q\}, k + B_i\} - B_{i+1}$ is the number of wasted steps initiated by $\mathcal{A}$, that is, all processes that ends before $Q$ but are computed based on incorrect prefix. 

Thus, the ultimate total number of tokens generated is the summation of $T_{B_i}$s:
\begin{align}
    \Sigma_{B_i\in B[:-1]}T_{B_i}
\end{align}

\textbf{Best case scenario} is that all steps generated by $\mathcal{A}$ matches those generated by $\mathcal{T}$, and therefore we do not have any ``wasted'' tokens, and then both $\mathcal{A}$ and $\mathcal{T}$ go through the agent generation plan. In this situation, $M_i = 0$ in the best case scenario for all $i$ corresponding to $B_i$ in $B$.
\begin{align}
    \Sigma_{0\leq i\leq n-1} (token(\mathcal{A}, s_i) + token(\mathcal{T}, s_i))
\end{align}

\textbf{Worst case scenario} is that none of the steps generated by $\mathcal{A}$ matches with those generated by $\mathcal{T}$. Additionally, each $\mathcal{T}$ process finishes after all $\mathcal{A}$ processes are completed, and the earliest called $\mathcal{T}$ process always finishes the last. Figure \ref{fig:worst_example} represents a partial example. Formally, the worst case scenario will occur under the condition which can be expressed as:
\begin{align}
   \forall B_i\in B, &\, end\_time(\mathcal{T}, s_{B_i+1}) \geq end\_time(\mathcal{A}, s_{B_{i+1}}) \text{ and }\\
   \forall B_i < l\leq B_{i+1}, &\, end\_time(\mathcal{T}, s_{B_i+1}) \geq end\_time(\mathcal{T}, s_{l}) \text{ and }\\
   \forall i\leq n-1, &\, a_i \text{ does not match } t_i 
\end{align}
In such a case, $Q = end\_time(B_i+1)$ and $M_i = k-1$ in the worst case scenario. Each $\mathcal{A}$ process $i$ will run for $(i\mod k)+1$ times (for example, the first process where $i = 0$ runs for 1 time, the $k$-th process where $i=k-1$ will run for $k$ times, and the $k+1$-th process where $i=k$ will run for 1 time), and each $\mathcal{T}$ process $i$ is run for $i$ times. Consequently, the total number of tokens generated in this worst-case scenario is:
\begin{align}
    \Sigma_{i=0}^{n-1} ((i\mod k)+1)*(token(\mathcal{A}, s_{i}) + token(\mathcal{T}, s_{i}))
\end{align}

\subsection{Rate Required}
This subsection focuses on analyzing the rate required to run the speculative planning algorithm, which is determined by the maximum number of concurrently running agent calls.

When not utilizing speculative planning, all agent calls are executed sequentially. Consequently, the required rate, which is the maximum number of concurrently running agent calls, is 1. When using speculative planning, we naturally have at least 2 concurrent calls: 1 for $\mathcal{A}$ and 1 for $\mathcal{T}$. But it can be more than 2, as shown in Figure \ref{fig:spec_plan} where we can have many $\mathcal{T}$ processes running at the same moment. To determine the maximum concurrent $\mathcal{C}$ processes, we identify the target agent process that overlaps with the most other target processes and add 1 for the additional approximation process. For all $\mathcal{T}_l$ processes for $B_{i} < l \leq B_{i+1}$, we find the $j$-th process $\mathcal{T}_j$ that overlaps with the most other $\mathcal{T}$ processes by:
\begin{align}
    \mathcal{T}_j = \max_{B_i < j\leq B_{i+1}}|\underbrace{\{l< n\mid start\_time(\mathcal{T}, s_l)\leq start\_time(\mathcal{T}, s_j) \leq end\_time(\mathcal{T}, s_l)\}}_{\text{count the number of target processes overlapping with process } j}|
\end{align}
We denote the number of overlapping processes to be $C_{T_i}$. Notice that we have a hyperparameter $k$ set up which controls the number of sequential $\mathcal{A}$ calls can be conducted without waiting for all corresponding $\mathcal{T}$ calls to be finished. Therefore, Note that $C_{T_i}$ is upper-bounded by $k$. Since the concurrent processes are the overlapping target process plus the approximation process, $\mathcal{C}_{B_i} = C_{T_i} + 1$ which is upper-bounded by $k+1$ between any consecutive $B_i$ and $B_{i+1}$.

Thus, the maximum concurrent $\mathcal{C}$ processes is the maximum of all $\mathcal{C}_{B_i}$:
\begin{align}
    \mathcal{C} = \max_{B_i\in B[:-1]}\mathcal{C}_{B_i}
\end{align}

\textbf{Best case scenario} is where there is exactly 2 concurrent processes running, 1 $\mathcal{A}$ process and 1 $\mathcal{T}$ process and there is no time overlap between any two $\mathcal{T}$ processes. This may only occur when for each step $s_i$, $time(\mathcal{T}, s_i)\leq time(\mathcal{A}, s_i)$.

\textbf{Worst case scenario} is when there is a sequence of steps $i$ to $i+k$ such that $\forall i< j\leq i+k, end\_time(\mathcal{T}, s_i) > start\_time(\mathcal{T}, s_j)$. In this case, there exists a time point where $k$ target processes are running concurrently, resulting in a total of $k+1$ concurrent processes.

\subsection{Simulation Experiment for Speculative Planning}
To elucidate the relationship between the performance of the Interactive Speculative Planning system and various hyperparameter configurations, we conducted three series of simulation experiments. Two experiments aimed to investigate the impact of different settings in speculative planning, specifically: (1) the choice of approximation agent $\mathcal{A}$, (2) the parameter $k$; and the third experiment investigates the impact of the number of user interruptions on overall latency. For the impact of $\mathcal{A}$, we examined $\mathcal{A}$'s accuracy relative to that of $\mathcal{T}$ (accuracy computed by treating $\mathcal{T}$'s result as ground truth), as well as $\mathcal{A}$'s computational speed. In the rest of the paper, we use $\mathcal{A}$'s accuracy to refer to the relative accuracy of $\mathcal{A}$ with respect to the result of $\mathcal{T}$.

For the simulation experiments, we set the following parameters unchanged: (1) the plan consists of 10 steps, (2) the generation speed of $\mathcal{T}$ is 8 seconds per action ($time(\mathcal{T}, s) = 8$) (3) for each step, $\mathcal{A}$ generates 10 tokens ($time(\mathcal{T}, s) = 10$) , (4) for each step, $\mathcal{T}$ generates 20 tokens ($time(\mathcal{T}, s) = 20$), and (5) for clarity, we set execution time to be 0 ($e(s) = 0$). 

The first series of experiments explores the impact of $\mathcal{A}$'s accuracy with respect to $\mathcal{T}$ and the hyperparameter of $k$ planning time and total tokens generated. We fix the speed of $\mathcal{A}$ ($time(\mathcal{A}, s) = 2$) to be 2 seconds per action and vary $\mathcal{A}$'s accuracy in $\{0.0, 0.1, 0.2, 0.3, 0.4, 0.5, 0.6, 0.7, 0.8, 0.9, 1.0\}$ and the hyperparameter of $k$ in $\{1, 2, 3, 4, 5, 6, 7, 8, 9, 10\}$.

\begin{figure}[!ht]
    \centering
    \includegraphics[scale=0.1]{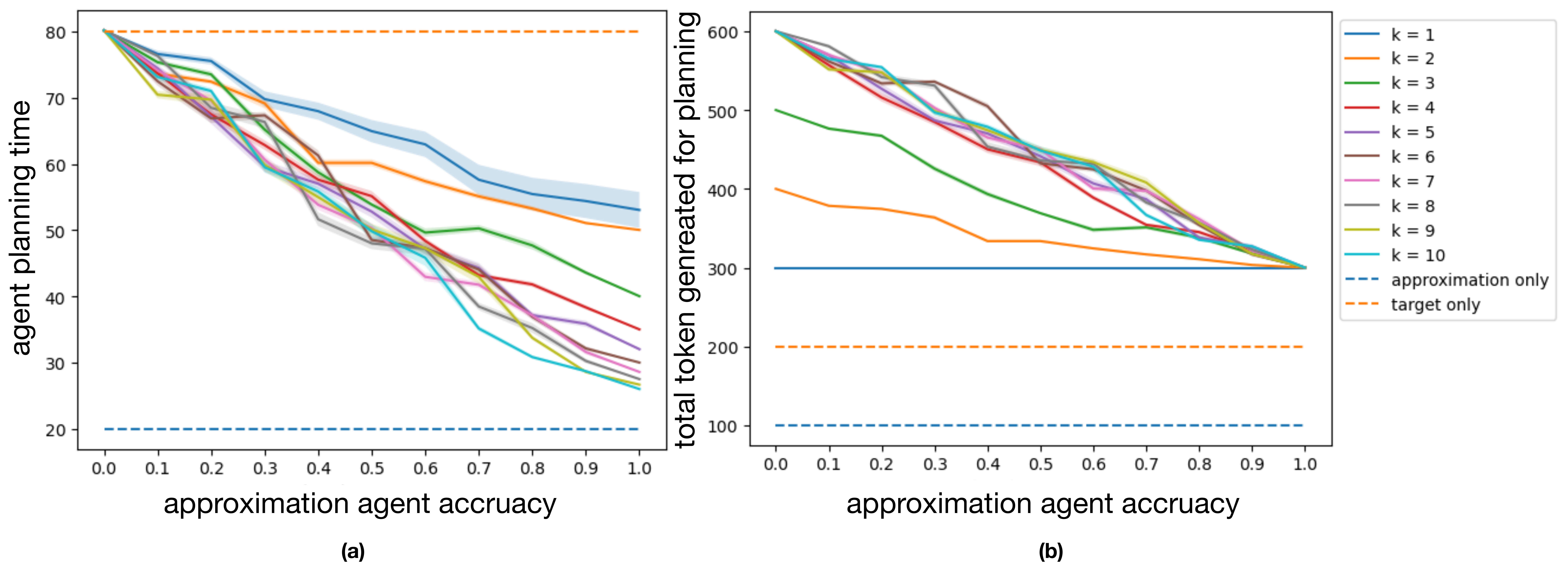}
    \caption{(a) Relationship between agent planning time, $\mathcal{A}$'s accuracy, and $k$ (b) Relationship between total token generated, $\mathcal{A}$'s accuracy, and $k$}
    \label{fig:k}
\end{figure}

Figure \ref{fig:k} (a) provides a visual representation of the impact of accuracy and the hyperparameter $k$ on the planning time. For each pair of accuracy and $k$, we run 10 experiments with different random seeds. The mean time and standard deviation of the average step are plotted in the figure, providing a comprehensive view of how these factors influence the planning time. We also present the time required for the agent planning when using $\mathcal{A}$ only and $\mathcal{T}$ only as in normal agent planning to show the lower bound and upper bound of speculative planning.

It is evident that higher accuracy in $\mathcal{A}$ results in shorter planning time. Very low $k$ (such as $k=1, 2, 3$) leads to slower agent planning, regardless of $\mathcal{A}$'s accuracy. For other $k$ values, as the accuracy increases, the impact of $k$ becomes more clear: higher $k$ leads to shorter agent planning time. However, when the accuracy is low, the impact is less clear.

Figure \ref{fig:k} (b) showcases the impact of accuracy and the hyperparameter $k$ on the total number of tokens generated during the planning process. In addition, the figure presents the tokens generated when using only $\mathcal{A}$ and when using only $\mathcal{T}$. There are two obvious trend: (1) higher accuracy in $\mathcal{A}$ generally results in a smaller number of tokens generated regardless of $k$ except in the trivial case when $k = 1$ (2) lower $k$ leads to a smaller number of tokens to be generated, especially when $k$ is small in the value range of $\{1, 2, 3, 4\}$; otherwise is impact is less clear.

\begin{figure}[!ht]
    \centering
    \includegraphics[scale=0.1]{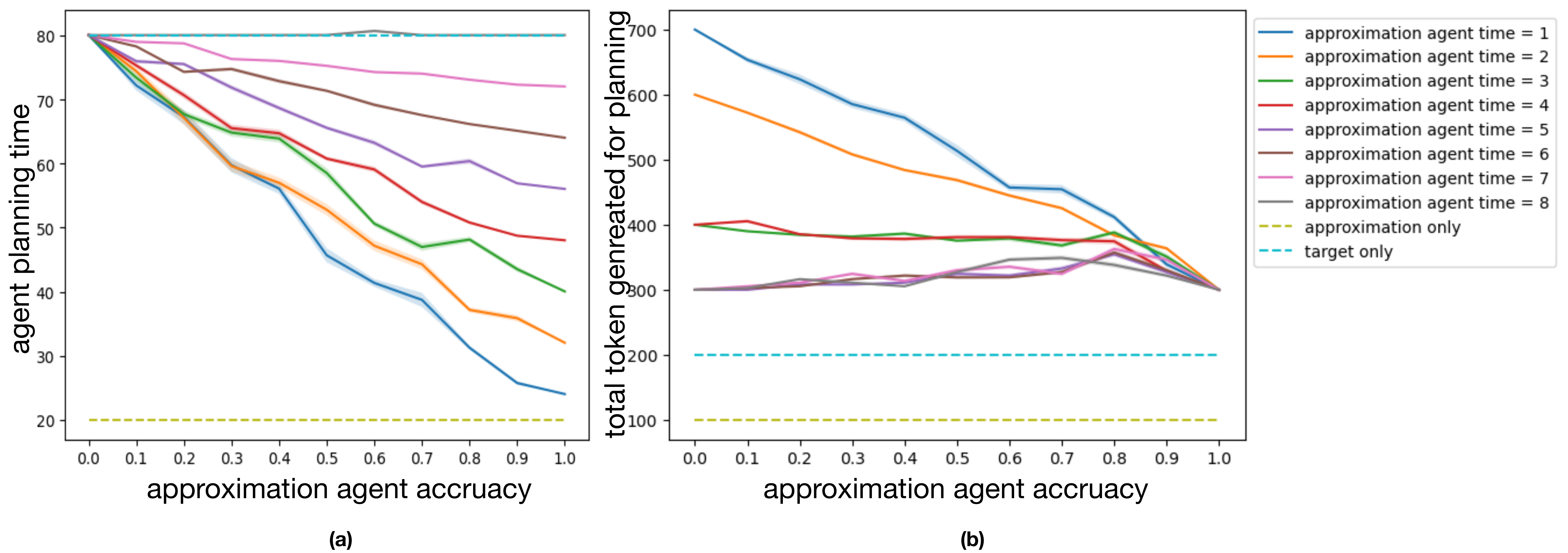}
    \caption{(a) Relationship between time and $\mathcal{A}$'s accuracy and speed (b) Relationship between time and $\mathcal{A}$'s accuracy and speed}
    \label{fig:ast}
\end{figure}

In the second series of experiments, we study the impact of $\mathcal{A}$'s speed and accuracy on planning time and generated tokens: We experiment on speed in different values: $\{1, 2, 3, 4, 5, 6, 7, 8\}$ and accuracy in values $\{0.0, 0.1, 0.2, 0.3, 0.4, 0.5, 0.6, 0.7, 0.8, 0.9, 1.0\}$. Here we set $k$ to be 5. Figure \ref{fig:ast} (a) demonstrates the change of planning time: (1) smaller speed values lead to smaller planning time, \emph{i.e.}, quicker planning, regardless of the accuracy of $\mathcal{A}$, and (2) better accuracy also leads to quicker planning, regardless of speed.

Figure \ref{fig:ast} (b) demonstrates the effect on total token generated. When the speed is very quick, higher accuracy monotonically reduces the total token generated. When the speed is around half of the speed of $\mathcal{T}$, accuracy does not have much impact on total tokens until it gets very high. When the speed is very slow (more than half of that of the target process), the total token generated first increases and then decreases as accuracy improves.

\begin{wrapfigure}{L}{0.5\textwidth}
\centering
\includegraphics[scale=0.23]{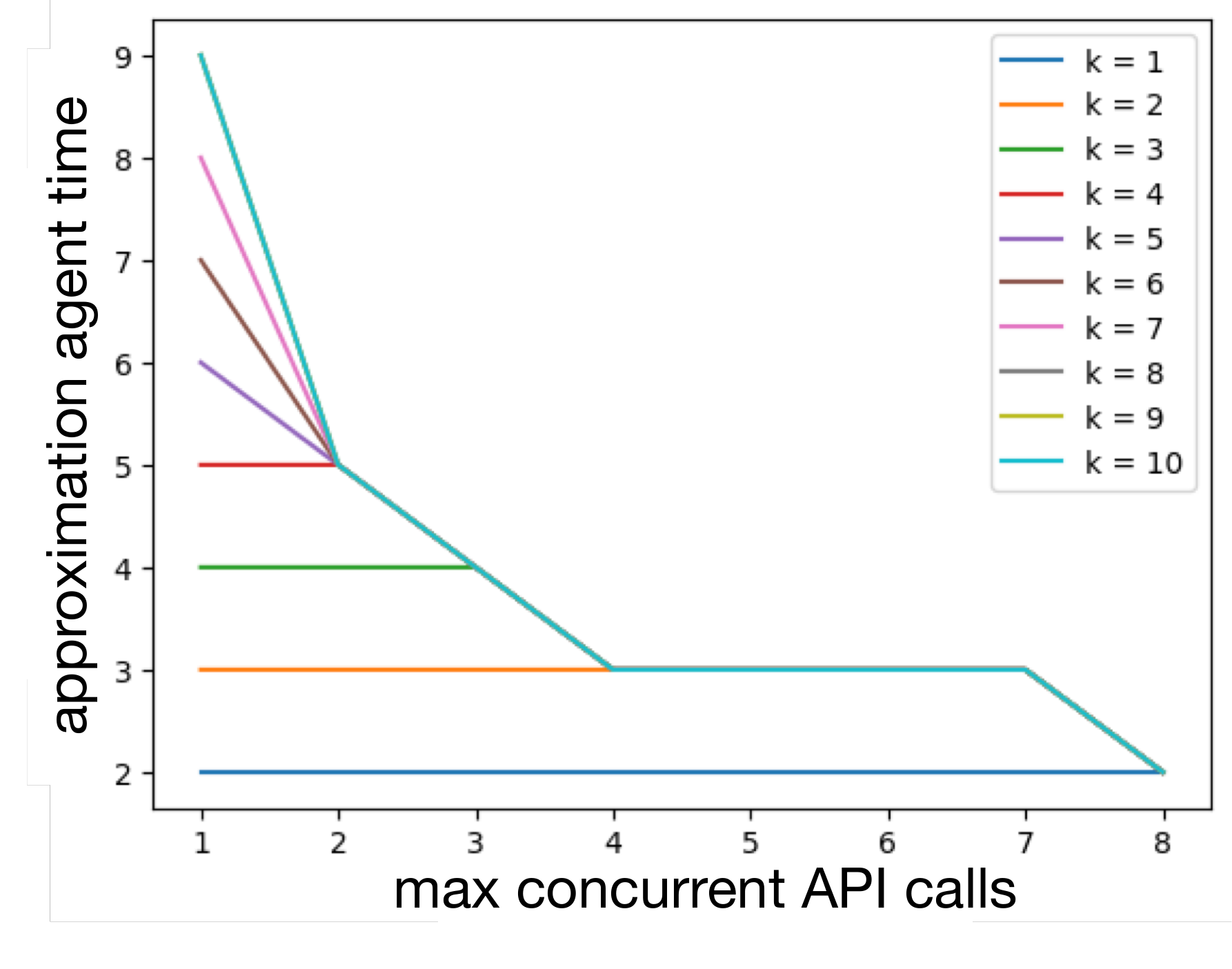}
\caption{Relationship between maximum concurrent rate required and $\mathcal{A}$'s speed and $k$}
\label{fig:rate}
\end{wrapfigure}

In the third series of experiments, we investigate the impact of user interruptions on time efficiency. We conduct simulations with varying interruption times. We assume that the user is actively monitoring the agent planning process and has a patience level between the speeds of the approximation agent $\mathcal{A}$ and the target agent $\mathcal{T}$, which assumption is made based on (1) $\mathcal{A}$ is designed to be an efficient agent (2) if the user's patience exceeds the speed of $\mathcal{T}$, no interruptions would occur.

For this simulation experiment, we set $k=n=10$ and the accuracy of $\mathcal{A}$ to be 0.5. The user is permitted to interrupt between 0 and 10 times. Each user interruption may occur randomly after waiting periods ranging from 1 to 5 seconds following the presentation of $\mathcal{A}$'s result.  For each number of user interruptions, we conduct 5 simulations.

\begin{wrapfigure}{L}{0.5\textwidth}
\centering
\includegraphics[scale=0.45]{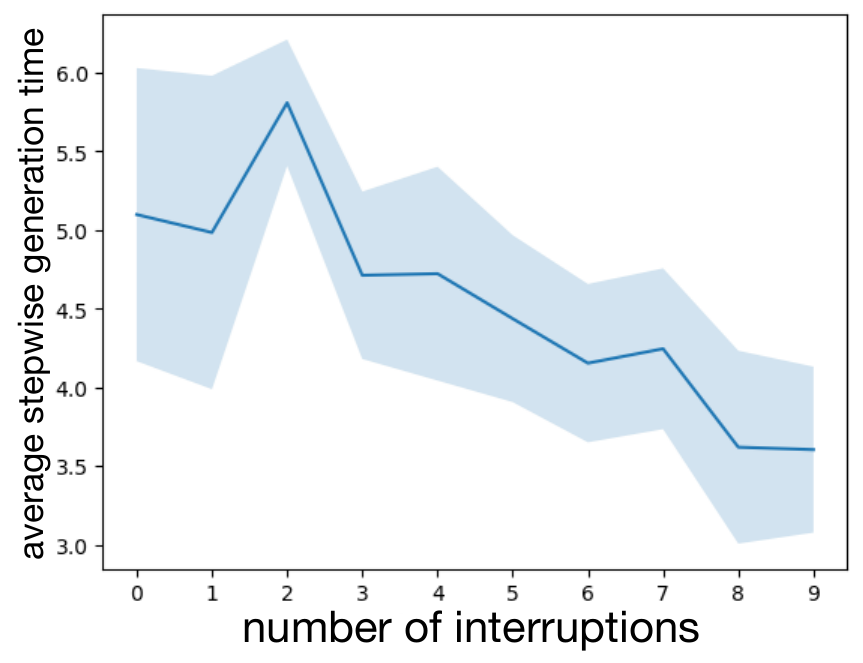}
\caption{Relationship between the number of user interruption simulation and stepwise generation time}
\label{fig:user_interruption_simulation}
\end{wrapfigure}

The results of this simulation are presented in Figure \ref{fig:user_interruption_simulation}, which displays the mean stepwise generation time along with the standard deviation. As anticipated, an increase in user interruptions reduces the overall latency of the system.

\section{Experiment}
\label{sec:experiment}
This section presents the results on two agent planning benchmarks: OpenAGI \citep{ge2024openagi} and TravelPlanner validation \citep{xie2024travelplanner}. For each benchmark, we attempt to implement four settings. We will briefly introduce the benchmarks and the settings in the below two subsections.

\subsection{Benchmarks}
OpenAGI is a benchmark designed for agent planning with complex tasks, built on computer vision and natural language processing-related tasks. Tools accessible to the agents include ``Sentiment Analysis''``Machine Translation''``Object Detection''``Visual Question Answering,,'' etc. An example task in the benchmark is ``Restore noisy, low-resolution, blurry, and grayscale images to regular images,'' whose solution is a sequence of tool usage: ``Image Super-resolution, Image Denoising, Image Deblurring, Colorization.'' This benchmark contains 117 multi-step tasks.

TravelPlanner is a planning benchmark that focuses on travel planning. It provides a rich sandbox environment, various tools for accessing nearly four million data records, and meticulously curated planning intents and reference plans. These plans also involve many constraints, including budget constraints, environmental constraints, etc. An example task is ``Please plan a travel itinerary for me. I'm departing from Cincinnati and heading to Norfolk for three days. The dates of travel are from March 10th to March 12th, 2022. I have a budget of \$1,400 for this trip.'' whose solution contains a sequence of actions such as ``FlightSearch[Cincinnati, Norfolk, 2023-03-12]'' where ``FlightSearch'' is the function name while ``Cincinnati, Norfolk, 2023-03-12'' are the natural language free-form parameters.

\subsection{Speculative Planning Settings}
To experiment with Interactive Speculative Planning in real-life scenarios, we demonstrate the performance using four different settings: four different combinations of the approximation agent $\mathcal{A}$ and the target agent $\mathcal{T}$.

\paragraph{Setting 1} $\mathcal{A}$ employs direct-generation-based planning with a GPT-4-turbo backbone, while $\mathcal{T}$ utilizes ReAct-based planning \citep{yao2022react} with the same backbone. For each step in the plan, $\mathcal{T}$ uses ReAct to first deliberate on the action and then generate it through two separate API calls, whereas $\mathcal{A}$ directly generates the action for that step.
\paragraph{Setting 2} $\mathcal{A}$ uses direct-generation-based planning with a GPT-4-turbo backbone, and $\mathcal{T}$ employs chain-of-thought (CoT)-based planning with the same backbone. For each step in the plan, $\mathcal{T}$ uses CoT to first reason and then generate the result in a single API call, while $\mathcal{A}$ directly generates the action for that step.
\paragraph{Setting 3} $\mathcal{A}$ uses CoT-based planning with a GPT-4-turbo backbone, and $\mathcal{T}$ system uses multi-agent-debate (MAD) including 2 agents with 2 rounds of discussion on every step of the plan with a GPT-4-turbo backbone. For each step in the plan, $\mathcal{T}$ system has two agents discuss with each other and finalize the action to take for the current step, and the while $\mathcal{A}$ uses CoT to first reason and then generate the result in a single API call for the step.
\paragraph{Setting 4} $\mathcal{A}$ uses direct-generation-based planning (DG) with a GPT-3.5-turbo backbone, and $\mathcal{T}$ uses direct-generation-based planning with a GPT-4-turbo backbone. In this setting, both $\mathcal{A}$ and $\mathcal{T}$ directly generate the result for each step. Notice that we cannot provide results for TravelPlanner in this setting, as direction generation using GPT-3.5-turbo fail to provide a valid action in many cases.

In all experiments, we set $k=4$. We utilized one OpenAI API for experiments under Settings 1, 2, and 4, and two OpenAI APIs (one API for each agent in the multi-agent system) for experiments under Setting 3. 

For the OpenAGI benchmark, given its limited action space, we used exact match to verify the correctness of the output generated by $\mathcal{A}$ against the output of $\mathcal{T}$. This ensured that the output of the speculative planning is the same as that of normal agent planning. For the TravelPlanner benchmark, which contains a much larger action space, each action is a combination of a function name from a fixed set and natural language free-form parameters. We verified the consistency between the output of $\mathcal{A}$ and the output of $\mathcal{T}$ based on an exact match of the function name and a soft match of the natural language parameters. The soft match is implemented by computing the Levenshtein distance: if the function name matched and the Levenshtein distance is smaller than 0.3, then the action is verified. As we leverage soft match to verify the output of $\mathcal{A}$, it is not guaranteed that the result from speculative planning remains the same as the result from normal agent planning and therefore we also provide the performance result of normal agent planning and speculative agent planning. Details in Appendix\ref{tp}.

\subsection{Evaluation Metrics}
In terms of latency, we report the average and minimum total generation time, as well as the stepwise generation time, for all planning tasks across each benchmark and experimental setting, compared with the normal planning setting. It is important to note that the total generation time heavily depends on the number of steps in the plan, which can be influenced by randomness. Therefore, we also report the stepwise generation time, which mitigates the effect of randomness related to the number of steps. To provide a comprehensive understanding of the algorithm, we also include metrics related to the total number of tokens generated during the process and the total API cost.

Therefore, in total there are 11 metrics: (1) total time (TT), (2) the minimum total time across the dataset (min-TT), (3) stepwise time (ST), (4) the minimum stepwise time across the dataset (min-ST), (5) Total tokens generated (TO), (6) the minimum total tokens generated across the dataset (min-TO), (7) stepwise tokens generated (SO), (8) the minimum stepwise tokens generated across the dataset (min-SO), (9) maximum concurrent API calls (MC) (10) the minimum maximum concurrent API calls across the dataset (min-MC), (11) the average total cost used to finish the plan (cost).

\subsection{Main Experiment Result}
\begin{table*}[h]
    \centering
    \resizebox{14cm}{!}{
    \begin{tabular}{rrr|rr|rr|rr}
            \toprule
           \textbf{Metrics}& \multicolumn{8}{c}{\bf Settings} \\
           &  Setting 1 & ReAct & Setting 2 & CoT & Setting 3 & MAD & Setting 4& DG  \\
         \midrule
           TT & 33.91$_{\pm 30.38}$& 43.63$_{\pm 25.39}$& 28.64$_{\pm 25.49}$& 39.96$_{\pm 27.25}$& 105.42$_{\pm 50.84}$& 182.70$_{\pm 421.49}$& 4.63$_{\pm 1.78}$ & 5.77$_{\pm 1.83}$ \\
           Min-TT & 6.80& 9.16& 3.53 & 8.60& 28.24& 50.89& 1.70 & 2.23 \\
           ST & 5.92$_{\pm 3.00}$& 8.69$_{\pm 2.75}$& 5.52$_{\pm 3.71}$& 7.98$_{\pm 2.72}$& 21.50$_{\pm 6.69}$& 34.84$_{\pm 58.94}$& 1.14$_{\pm 0.25}$ & 1.49$_{\pm 0.43}$ \\
           Min-ST & 2.33& 4.41& 0.50 & 3.81& 11.70& 19.21& 0.75 & 1.03 \\
           TO & 1920$_{\pm 879.79}$& 1812.89$_{\pm 832.30}$ & 1770.61$_{\pm 1010.44}$& 1397.90$_{\pm 794.55}$& 6781.43$_{\pm 3159.84}$& 4075.4$_{\pm 1603.54}$& 107.05$_{\pm 38.76}$ & 40.13$_{\pm 13.39}$ \\
           Min-TO & 760 & 652 & 455 & 352 & 1754& 1441& 47 & 17 \\
           SO & 288.72$_{\pm 65.29}$& 266$_{\pm 44.37}$ & 281.92$_{\pm 88.77}$ & 229.45$_{\pm 44.23}$ & 1385$_{\pm 391.77}$& 836.65$_{\pm 112.06}$& 26.47$_{\pm 5.06}$ & 10.14$_{\pm 1.98}$ \\
           Min-SO & 190.00 & 166.58 & 143.83 & 162.8 & 877& 558.33& 19.25 & 8.5 \\
           MC & 4.66$_{\pm 0.59}$& 1$_{\pm 0.00}$ & 4.49$_{\pm 0.82}$& 1$_{\pm 0.00}$ & 4.53$_{\pm 0.56}$& 1$_{\pm 0.00}$ & 4.05$_{\pm 0.21}$ & 1$_{\pm 0.00}$ \\
           Min-MC & 3 & 1 & 3 & 1 & 3& 1 & 4 & 1 \\  
           cost & 0.122$_{\pm 0.072}$& 0.0713$_{\pm 0.026}$& 0.074$_{\pm 0.040}$&0.044$_{\pm 0.018}$&0.2973$_{\pm 0.1387}$&0.2160$_{\pm 0.0795}$& 0.0012$_{\pm 0.0011}$& 0.0012$_{\pm 0.0004}$ \\
           \bottomrule
    \end{tabular}
    }
    \caption{Main experiment results on OpenAGI benchmark.}
    \label{tab:openagi}
\end{table*}

Table.\ref{tab:openagi} presents the results on OpenAGI dataset. In Setting 1 where $\mathcal{T}$ utilizes ReAct, we cut the total running time on average by 22.27\% percentage and the stepwise running time on average by about 31.87\%. In Setting 2 where $\mathcal{T}$ utilizes CoT, we cut the total running time on average by 28.32\% and the stepwise running time on average by about 30.83\%. In Setting 4 where both $\mathcal{A}$ and $\mathcal{T}$ uses direct generation but with different backbone models, we cut the total running time on average about 20.37\% percentage and the stepwise running time on average by about 23.50\%. \textit{Setting 3 includes a very slow $\mathcal{T}$ using a multi-agent debate; we obtain the largest efficiency improvement: this setting can cut the total time by 42.30\% and the stepwise running time on average by 38.29\%.}

Table.\ref{tab:travelplanner} presents the results on TravelPlanner validation dataset. Similar to the experiment on OpenAGI dataset, we can find noticeable latency improvement when using speculative planning: in Setting 1, the average latency on total generation time has decreased for 21.43\% while the stepwise generation time has decreased for 29.52\%; Setting 2 has decreased average total time by 19.18\% and the stepwise generation time by 32.53\%; Setting 3 has decreased average total time by 25.46\% and the stepwise generation time by 31.69\%.
\begin{table*}[h]
    \centering
    \resizebox{14cm}{!}{
    \begin{tabular}{rrr|rr|rrll}
            \toprule
          \multirow{2}{*}{\bf Metrics}& \multicolumn{6}{c}{\bf Settings} & \multicolumn{2}{c}{}\\
           &  Setting 1 & ReAct & Setting 2 & CoT & Setting 3 & MAD & Setting 4& DG\\
         \midrule
           TT & 137.33$_{\pm 66.39}$& 176.28$_{\pm 77.18}$& 98.09$_{\pm 45.02}$ & 121.37$_{\pm 32.18}$ & 568.10$_{\pm 292.99}$ & 733.12$_{\pm 290.51}$  & -& -\\
           Min-TT & 40.78& 55.18& 29.22 & 42.09 & 149.00 & 127.59 & -& -\\
           ST & 11.16$_{\pm 5.49}$& 14.13$_{\pm 3.61}$& 10.71$_{\pm 5.50}$ & 12.75$_{\pm 4.33}$ & 27.53$_{\pm 8.67}$ & 40.03$_{\pm 8.74}$ & -& -\\
           Min-ST & 4.53& 7.04& 2.65 & 4.92& 12.33 & 23.06  & -& -\\
           TO & 3751.94$_{\pm 853.86}$& 2460.95$_{\pm 332.07}$ & 3082$_{\pm 235.09}$ & 2002.93$_{\pm 276.54}$ & 12353.84$_{\pm 5872.86}$ & 8976.39$_{\pm 5371.31}$ & -& -\\
           Min-TO & 1389& 1762 & 833 & 1329 & 3443 & 2049 & -& -\\
           SO & 298.84$_{\pm 128.97}$ & 246.13$_{\pm 56.34}$ & 220.79$_{\pm 56.19}$ & 197.08$_{\pm 87.68}$ & 733.18$_{\pm 477.72}$ & 591.65$_{\pm 467.82}$  & -& -\\
           Min-SO & 128.13 & 108.30 & 85.42 & 68.06 & 189.00 & 186.27  & -& -\\
           MC & 5$_{\pm 0.00}$& 1$_{\pm 0.00}$ & 5$_{\pm 0.00}$ & 1$_{\pm 0.00}$ & 5.00$_{\pm 0.00}$ & 1$_{\pm 0.00}$  & -& -\\
           Min-MC & 5& 1 & 5& 1 & 5 & 1 & -& -\\  
           cost & 0.1583$_{\pm 0.0367}$& 0.1038$_{\pm 0.0033}$& 0.1393$_{\pm0.0241}$&0.0874$_{\pm0.0125}$& 0.5941$_{\pm 0.2871}$& 0.3990$_{\pm 0.2309}$ & -& -\\
           \bottomrule
    \end{tabular}
    }
    \caption{Main experiment results on TravelPlanner benchmark.}
    \label{tab:travelplanner}
\end{table*}

Notice that our experiments in this section do not indicate the upper bound of efficiency improvement in the two datasets but rather the performance based on the current settings.

\subsection{Analysis of Latency Improvement Breakdown}



Having observed the average latency improvement for the two datasets across the four settings, we now turn our attention to a more granular analysis of the latency improvement based on the accuracy of the approximation agent $\mathcal{A}$. Specifically, we aim to examine how much time is saved given a specific level of $\mathcal{A}$'s accuracy. This analysis will allow us to identify the sources of time savings and determine which datapoints, at which levels of accuracy, contribute to latency improvement and which do not. 

Thus, in this section, for each dataset and each setting, we present two figures: one displaying the distribution of datapoints with different levels of accuracy, and the other displaying the average stepwise latency improvement proportion for all levels of accuracy.

\begin{figure}[!ht]
    \centering
    \includegraphics[scale=0.4]{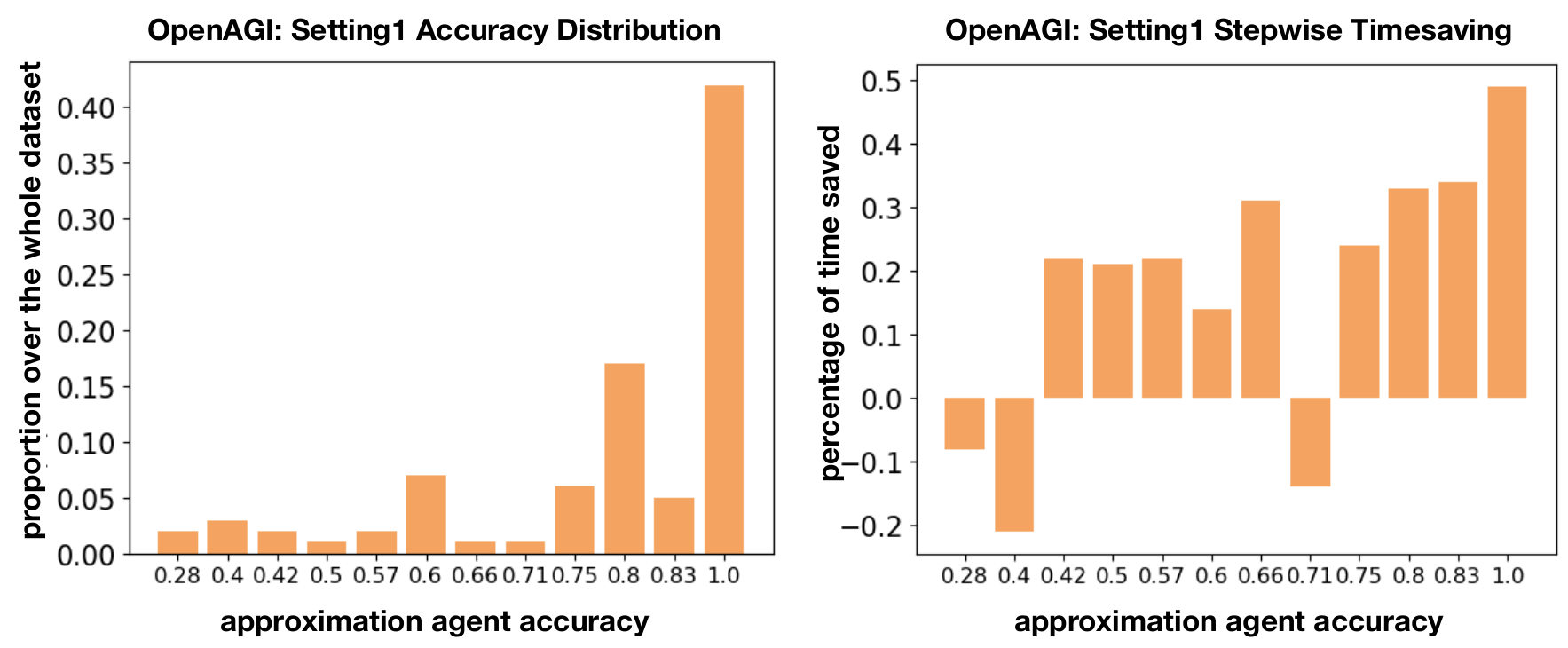}
    \caption{Distribution of $\mathcal{A}$'s accuracy in Setting 1 and corresponding latency improvement}
    \label{fig:openagi_accuracy_distribution_cot}
\end{figure}

\begin{figure}[!ht]
    \centering
    \includegraphics[scale=0.4]{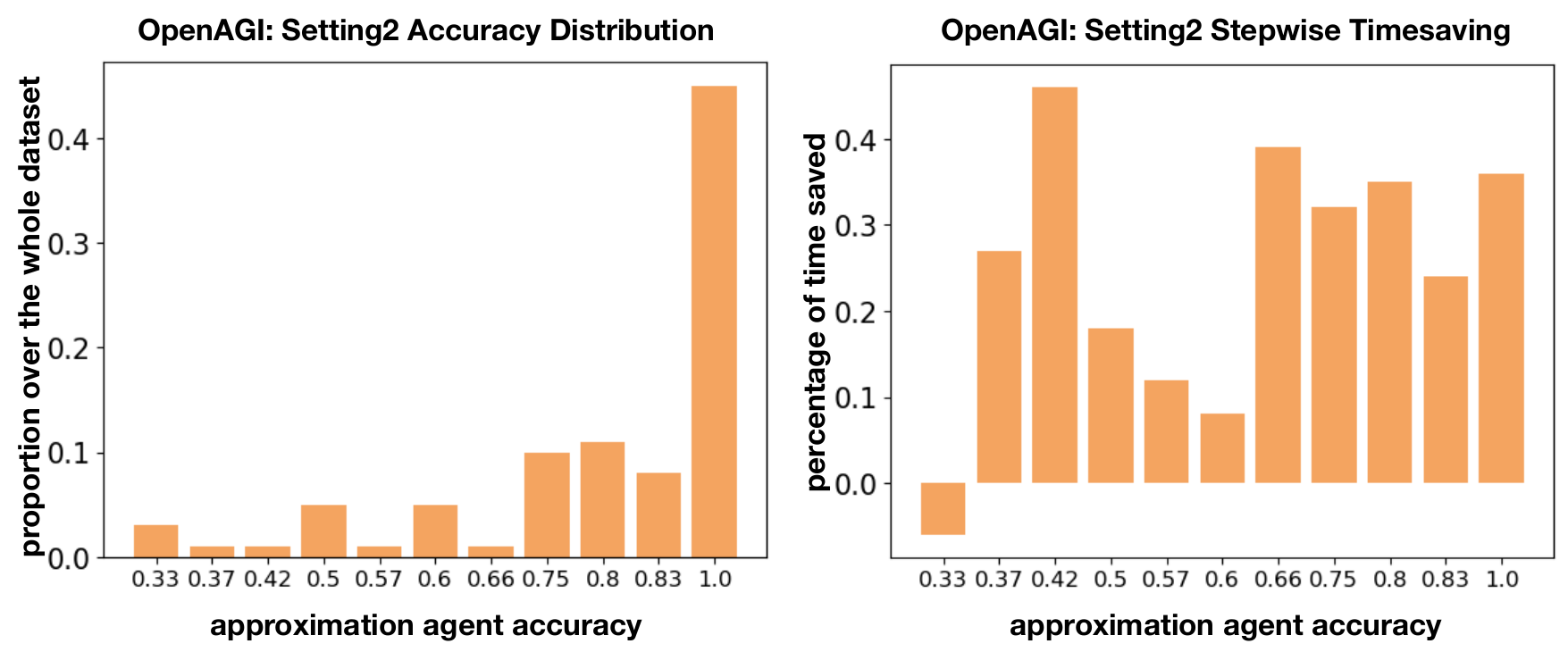}
    \caption{Distribution of $\mathcal{A}$'s accuracy in Setting 2 and corresponding latency improvement}
    \label{fig:openagi_timesaving_distribution_cot}
\end{figure}

\begin{figure}[!ht]
    \centering
    \includegraphics[scale=0.4]{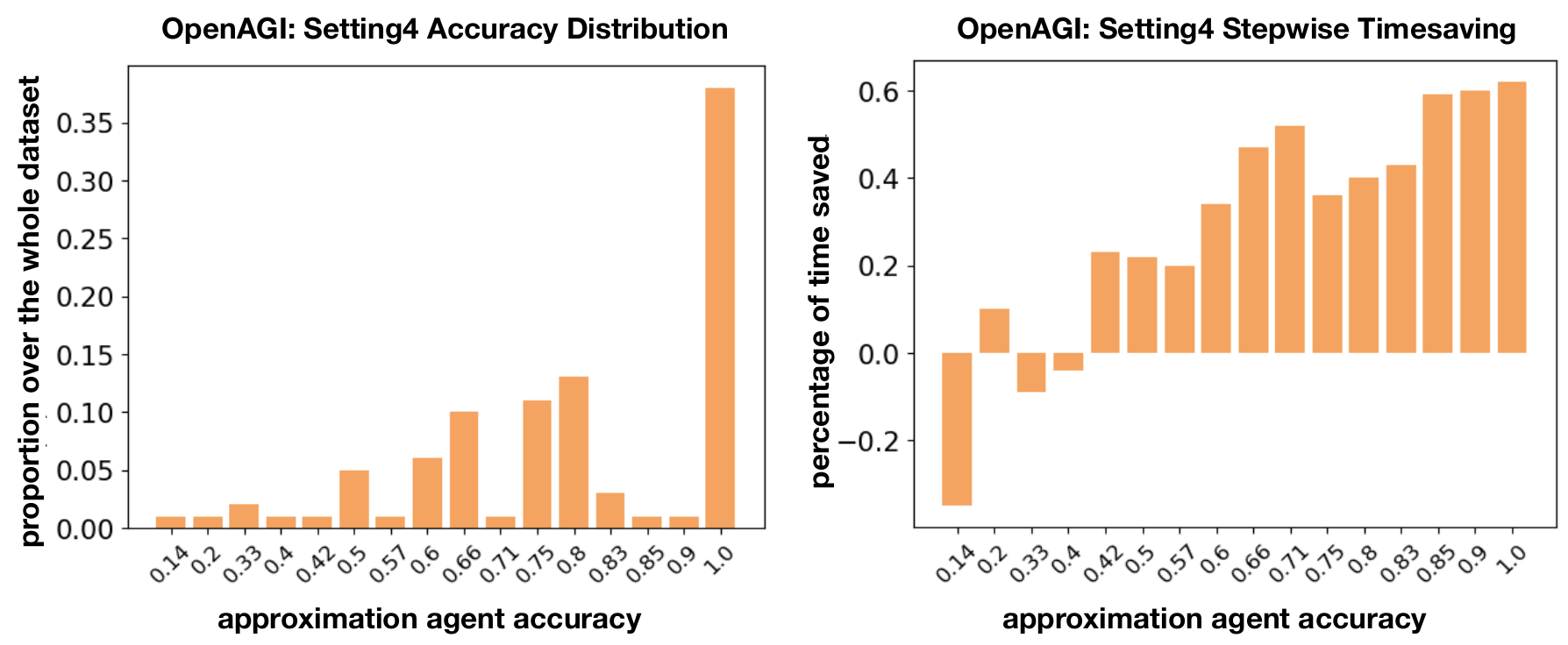}
    \caption{Distribution of $\mathcal{A}$'s accuracy in Setting 3 and corresponding latency improvement}
    \label{fig:openagi_timesaving_distribution_mad}
\end{figure}

\begin{figure}[!ht]
    \centering
    \includegraphics[scale=0.4]{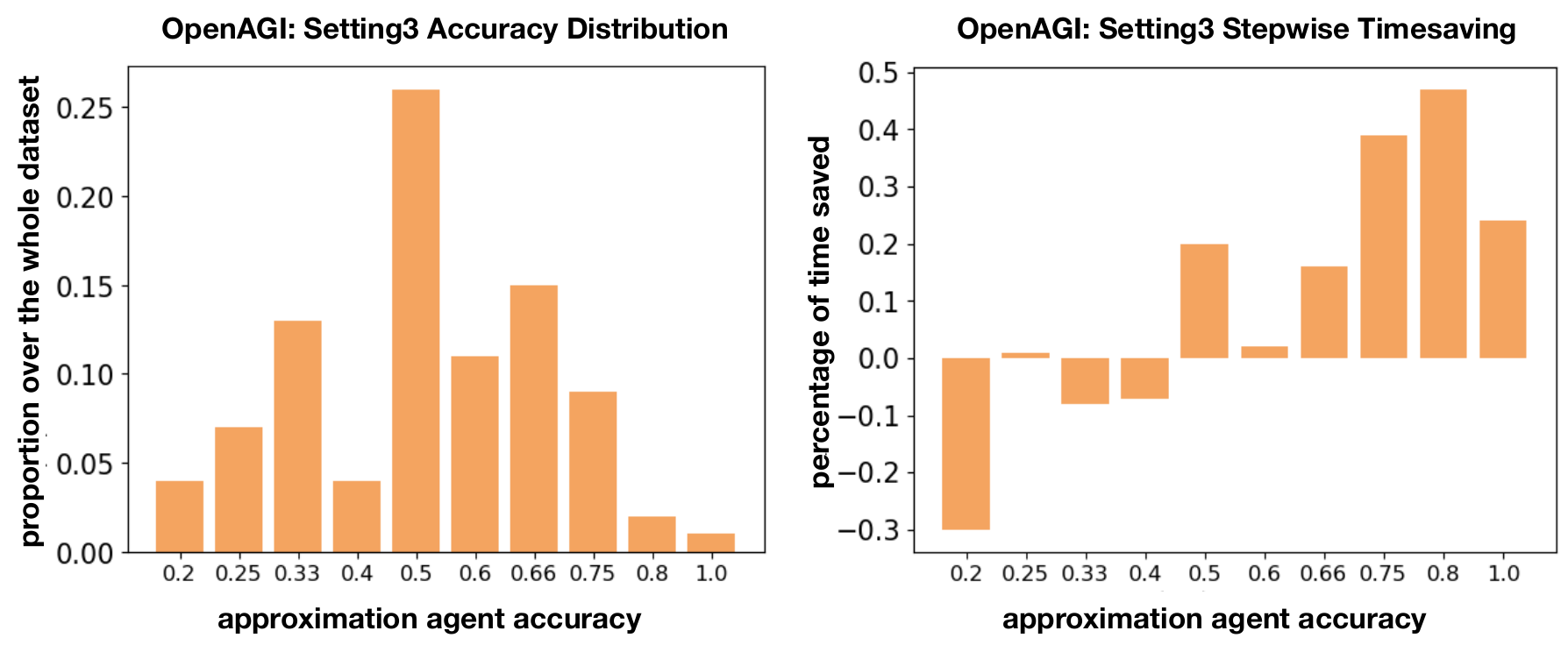}
    \caption{Distribution of $\mathcal{A}$'s accuracy in Setting 4 and corresponding latency improvement}
    \label{fig:openagi_timesaving_distribution_direct}
\end{figure}


Figure \ref{fig:openagi_accuracy_distribution_cot}, \ref{fig:openagi_timesaving_distribution_cot},  \ref{fig:openagi_timesaving_distribution_direct} and \ref{fig:openagi_timesaving_distribution_mad} demonstrate the breaking-down results on OpenAGI dataset. Notably, Setting 1, 2, and 3 contain a significant proportion of datapoints that exhibit a perfect accuracy of $\mathcal{A}$. Such datapoints also correspond to the largest latency improvement. Nevertheless, even with lower accuracy approximations, a substantial reduction in stepwise generation time can also be observed. This trend is consistent across all settings. And notice that in Setting 3, stepwise time saving proportion can achieve almost 60\% when $\mathcal{A}$'s accuracy achieves higher then 80\%. 

\begin{figure}[!ht]
    \centering
    \includegraphics[scale=0.4]{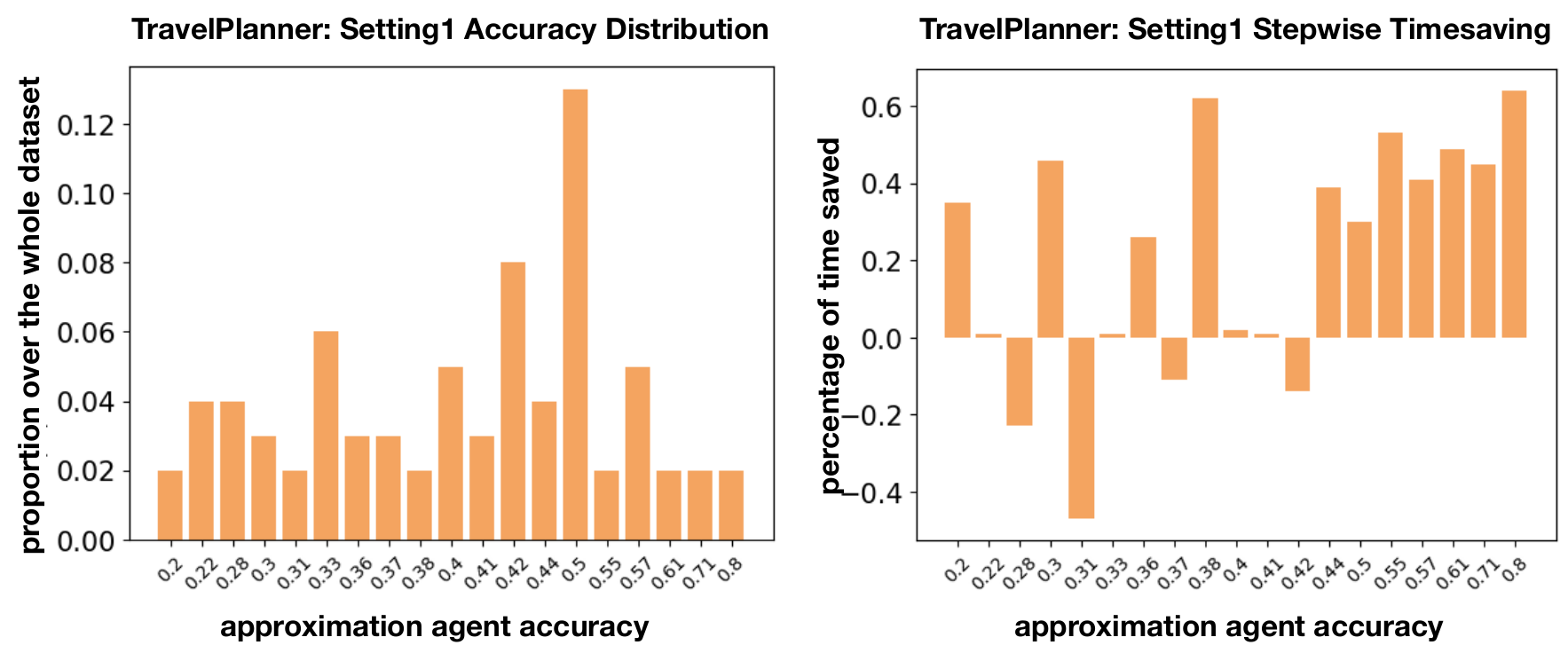}
    \caption{Distribution of $\mathcal{A}$'s accuracy in Setting 1 and corresponding latency improvement}
    \label{fig:tp_accuracy_distribution_react}
\end{figure}

\begin{figure}[!ht]
    \centering
    \includegraphics[scale=0.4]{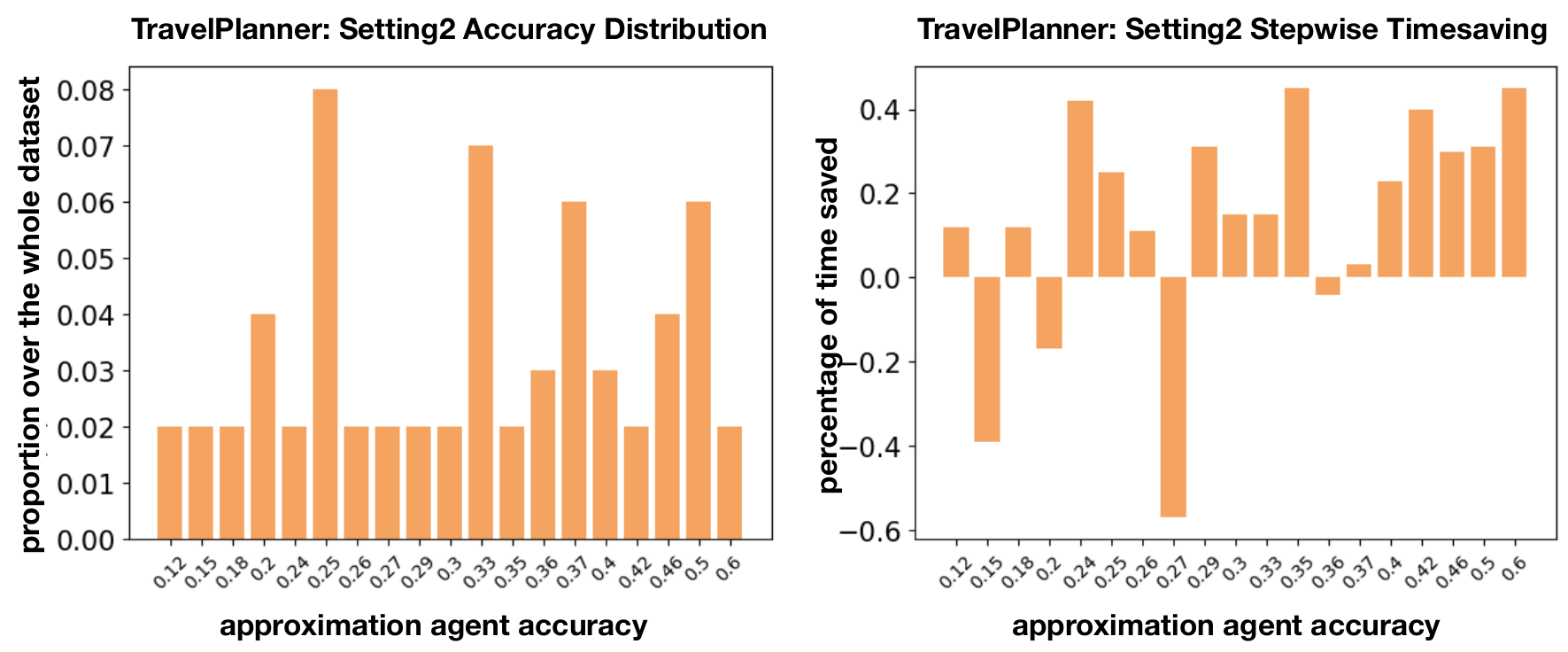}
    \caption{Distribution of $\mathcal{A}$'s accuracy in Setting 2 and corresponding latency improvement}
    \label{fig:tp_timesaving_distribution_cot}
\end{figure}

\begin{figure}[!ht]
    \centering
    \includegraphics[scale=0.4]{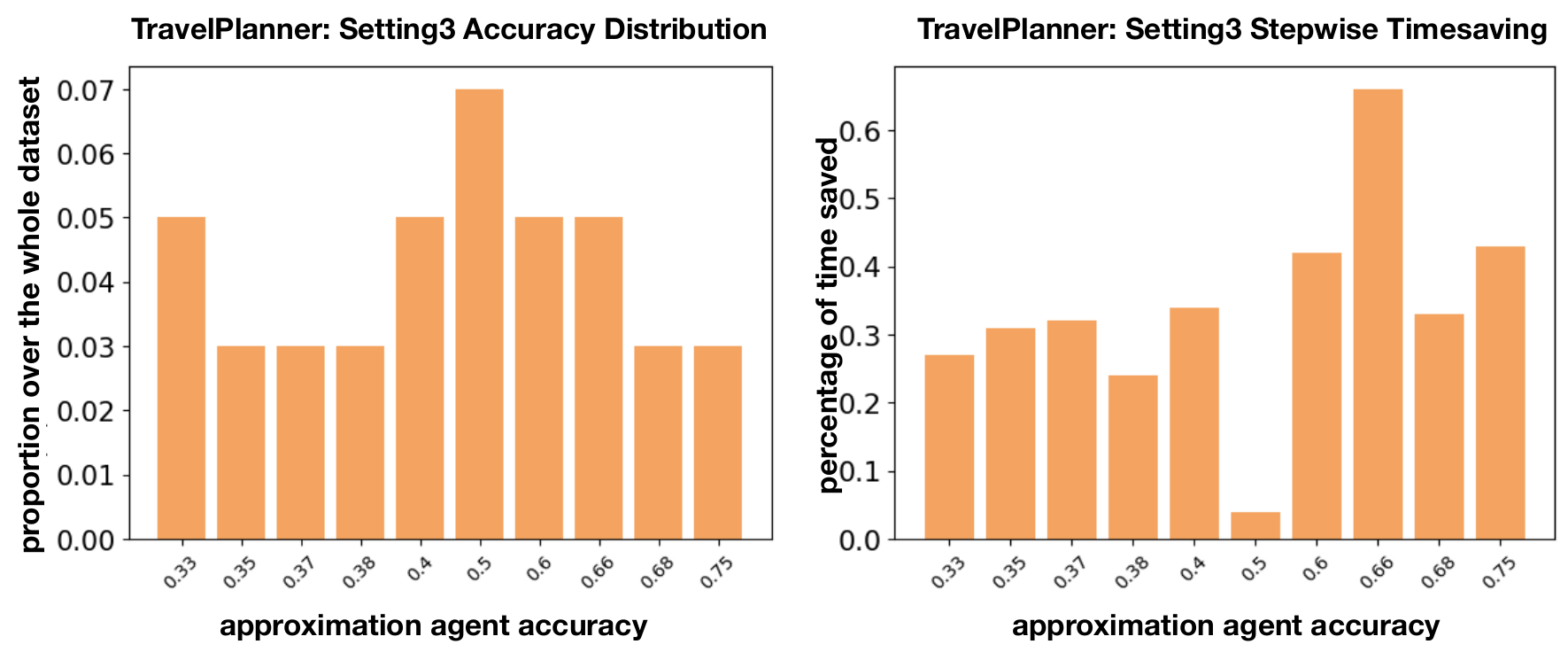}
    \caption{Distribution of $\mathcal{A}$'s accuracy in Setting 3 and corresponding latency improvement}
    \label{fig:tp_timesaving_distribution_direct}
\end{figure}

Figure \ref{fig:tp_accuracy_distribution_react}, \ref{fig:tp_timesaving_distribution_cot}, and \ref{fig:tp_timesaving_distribution_direct} present the results for the TravelPlanner dataset. In TravelPlanner, the distribution of datapoints based on accuracy is flatter (we only show accuracy levels where the proportion of datapoints exceeds 2\% to avoid excessive randomness). In all three settings, the latency improvement can achieve more than 40\% when the accuracy exceeds approximately 0.4.

However, it is important to note that in almost all settings for both datasets, there are data points exhibiting ``negative'' latency improvement, i.e., longer stepwise running times when applying speculative planning compared to normal agent planning. Most of these cases occur when the accuracy of $\mathcal{A}$ is relatively low, a scenario in which latency efficiency analysis suggests limited latency improvement but no worse than normal agent planning, contrary to what we observe. This discrepancy can be attributed to two assumptions in the theoretical analysis: (1) The speeds of $\mathcal{A}$ and $\mathcal{T}$ are constant across different runs on the same data points. However, in actual usage, there is significant randomness involved due to the number of tokens generated in each step, causing variations in the speed of both $\mathcal{A}$ and $\mathcal{T}$ even for the same data point. (2) The speeds of $\mathcal{A}$ and $\mathcal{T}$ are not affected by multiple concurrent queries. In practice, $\mathcal{T}$ runs in parallel, meaning the API for $\mathcal{T}$ must process multiple concurrent queries, which may also slow down the overall speed for each individual call of $\mathcal{T}$.

\subsection{Analysis of User Interaction}
One of the motivations behind Interactive Speculative Planning is users' patience. Numerous studies \citep{horvitz1999principles, barron2004graphical, simpson2007impact, carr1992effects} have demonstrated the physiological and psychological impacts of interaction delays on human-computer interaction. Therefore, we aim to quantitatively study how speculative planning enhances user experience by analyzing the frequency with which a user may become impatient and desire to interact with or interrupt the system.

\begin{figure}[!ht]
    \centering
    \includegraphics[scale=0.35]{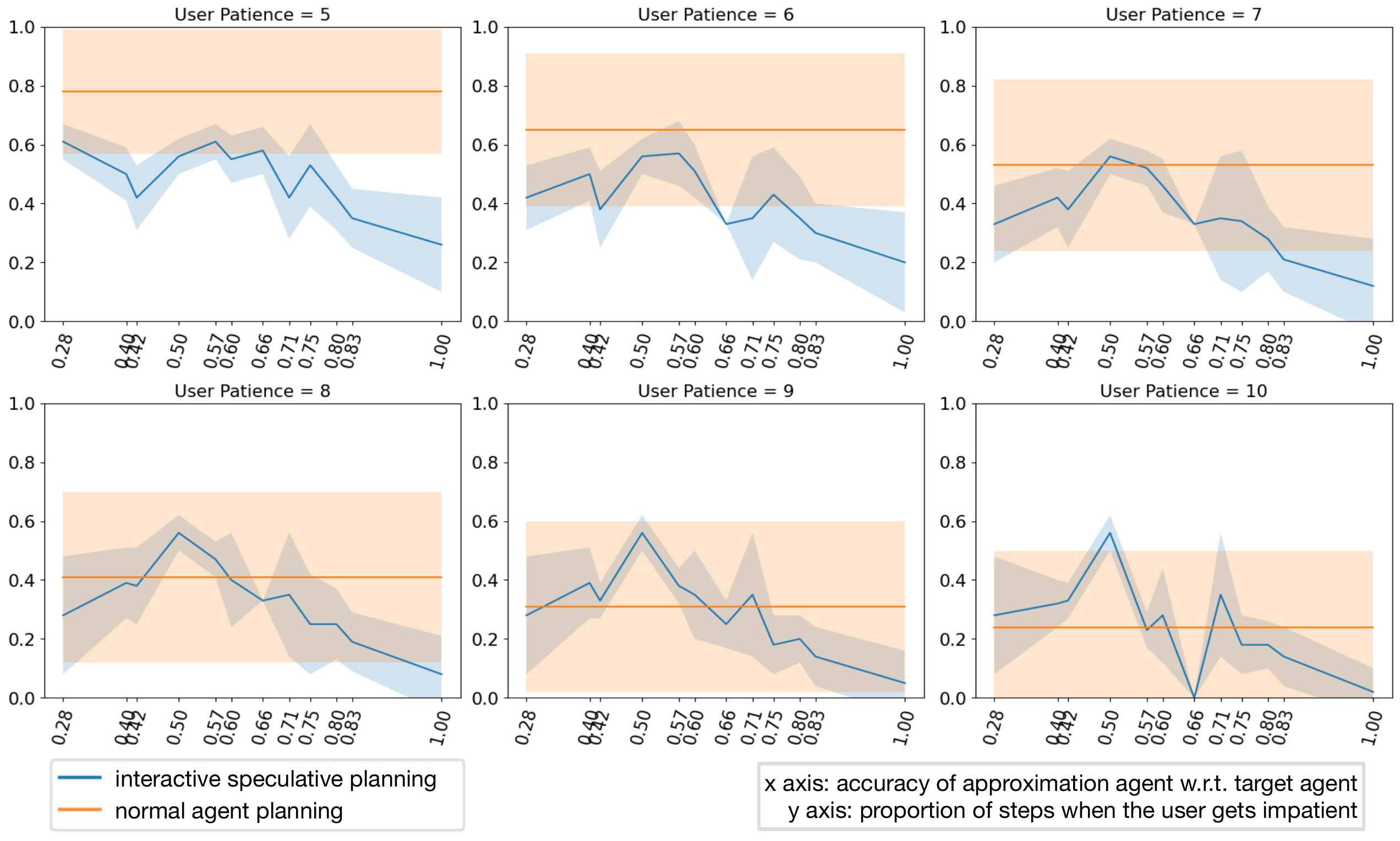}
    \caption{Number of Potential User Interruptions with Setting 1 on OpenAGI dataset and corresponding normal agent planning}
    \label{fig:openagi_react_user_interruption}
\end{figure}

For the quantitative study, we use the OpenAGI dataset with Setting 1, 2, and 3 as examples\footnote{We do not adopt Setting 4 here as the average stepwise time for both normal agent planning and speculative planning is too short.}. We collect statistics, including the mean and variance, on the number of user interruptions that may occur due to impatience by simulating users with different impatience thresholds. For Settings 1 and 2, we simulate users with impatience thresholds of 5, 6, 7, 8, 9, and 10 seconds. For Setting 3, which takes a much longer time to run, we simulate users with impatience thresholds of 11, 13, 17, 19, and 21 seconds. We also collect statistics for normal agent planning for comparison. For each setting, we provide a series of six figures. In each figure, the x-axis represents $\mathcal{A}$'s accuracy, and the y-axis represents the proportion of steps for which the user may become impatient. Each figure demonstrates the number of times the user may become impatient and interact with the system under Interactive Speculative Planning and normal agent planning, with respect to groups of data points with different levels of $\mathcal{A}$'s accuracy.

\begin{figure}[!ht]
    \centering
    \includegraphics[scale=0.35]{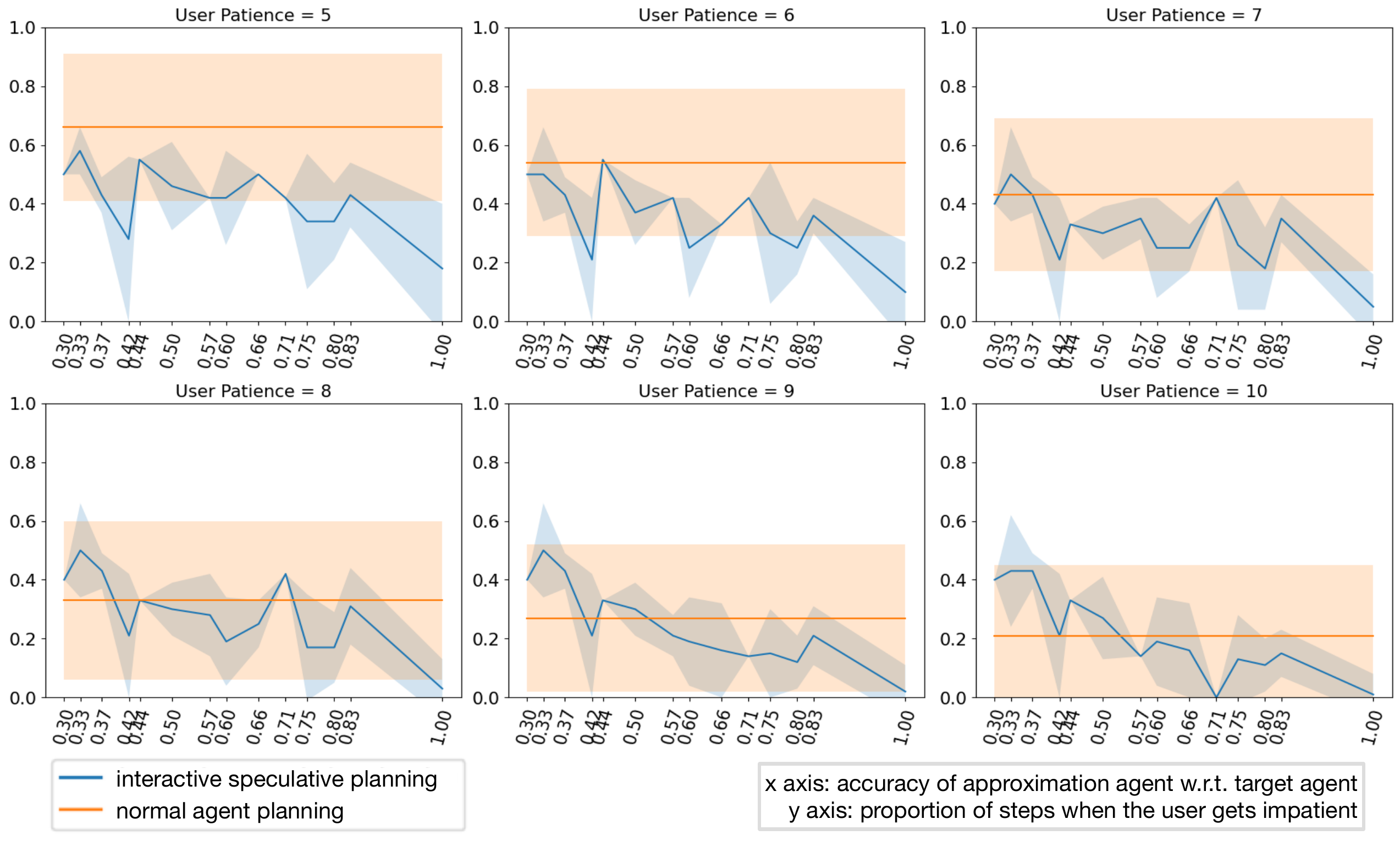}
    \caption{Number of Potential User Interruptions with Setting 2 on OpenAGI dataset and corresponding normal agent planning}
    \label{fig:openagi_cot_user_interruption}
\end{figure}

\begin{figure}[!ht]
    \centering
    \includegraphics[scale=0.35]{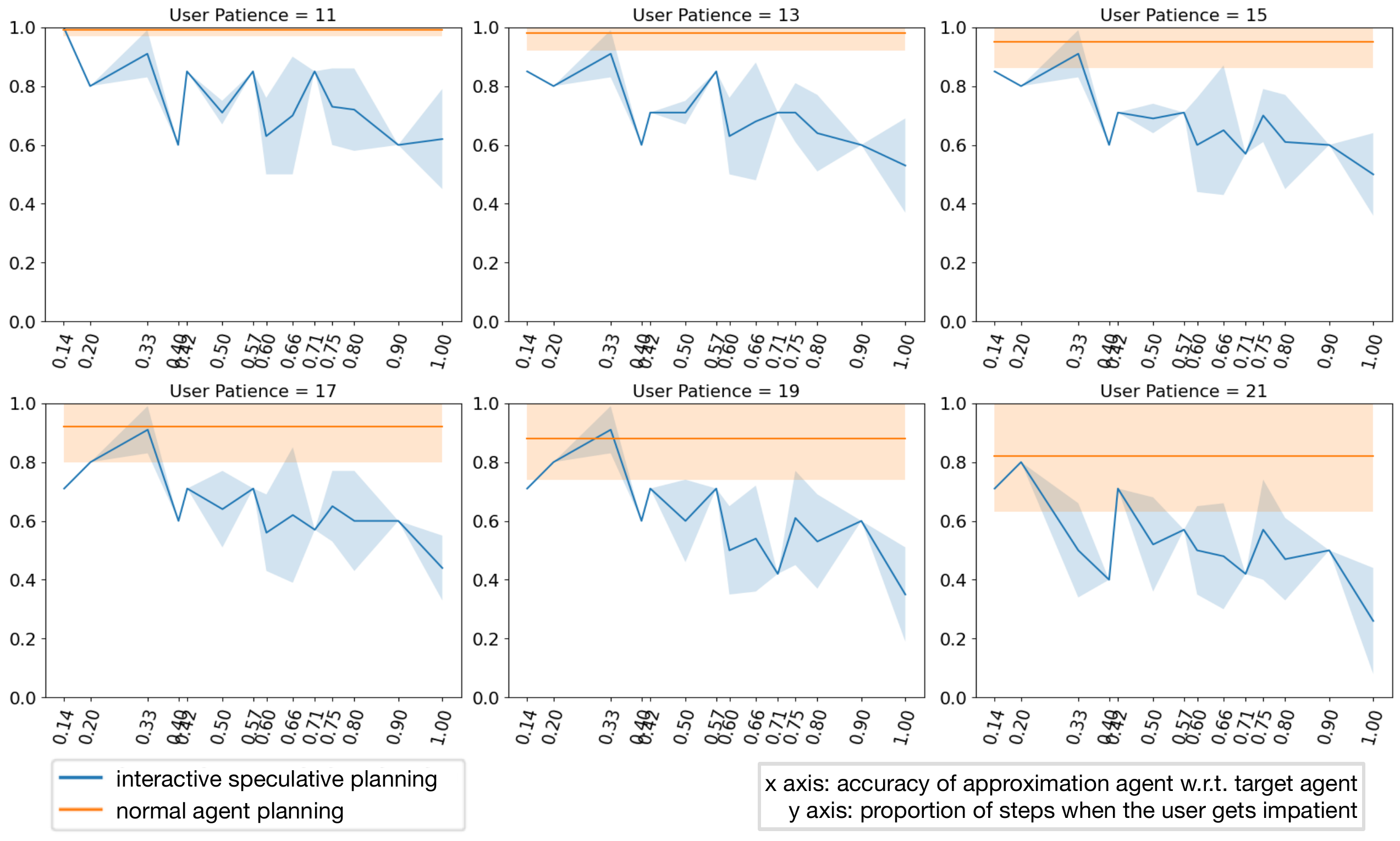}
    \caption{Number of Potential User Interruptions with Setting 3 on OpenAGI dataset and corresponding normal agent planning}
    \label{fig:openagi_mad_user_interruption}
\end{figure}

Figures \ref{fig:openagi_react_user_interruption}, \ref{fig:openagi_cot_user_interruption}, and \ref{fig:openagi_mad_user_interruption} represent the results for Settings 1, 2, and 3, respectively. As expected, Interactive Speculative Planning exhibits more observable differences for more impatient users.

\section{Limitations and Future Directions}
\label{sec:limitation}
Interactive Speculative Planning represents the first attempt at co-designing an efficient agent system alongside an active user interface. Consequently, it is imperfect in many aspects, and there are numerous future directions to be explored:
\paragraph{Spectre} 
Spectre \citep{mcilroy2019spectre, kocher2020spectre} refers to vulnerabilities and attacks involved in speculative execution \citep{kocher2020spectre, gabbay1996speculative, nightingale2005speculative}, a hardware feature that improves processor performance by predicting a program's future execution and executing instructions ahead of the current instruction pointer. This concept is analogous to speculative planning. However, speculative execution is known to create security vulnerabilities that allow attackers to access sensitive data. In our algorithm, the execution of $\mathcal{A}$'s steps that are unverified or inconsistent with $\mathcal{T}$'s steps also introduces vulnerabilities. Studies \citep{hua2024trustagent} have shown that smaller and weaker models are more prone to unsafe and untrustworthy actions. Therefore, the applicability of Interactive Speculative Planning in its current form should be constrained to non-high-stakes areas. 

To address the security vulnerabilities associated with speculative execution in Interactive Speculative Planning, several solutions can be implemented:

\begin{enumerate}
    \item Human-in-the-Loop Verification: Incorporate human-in-the-loop mechanisms to double-check the security of $\mathcal{A}$'s actions before execution. This approach leverages human oversight to ensure that potentially unsafe actions are identified and mitigated before they can cause harm.
    \item Isolated Execution Environments: Execute $\mathcal{A}$'s actions in isolated environments, such as Docker containers. This isolation ensures that any potentially malicious or untrustworthy actions are contained and do not affect the broader system or access sensitive data.
\end{enumerate}

By implementing these solutions, we can enhance the security of Interactive Speculative Planning, making it more robust and suitable for a wider range of applications, including those in high-stakes environments.

\paragraph{Effectiveness of step-by-step comparison}
In the current version of speculative planning, we employ a simple and straightforward method to determine whether an action proposed by $\mathcal{A}$ can be accepted: exact match. However, it is widely recognized that completing a task often involves multiple different paths and plans, and ``difference'' does not necessarily imply ``incorrect.'' There are two types of differences to consider: (1) different surface strings may refer to the same step, and (2) different steps may refer to two acceptable paths for the planning. Therefore, a step-by-step exact match judgment for whether $\mathcal{A}$'s output is accepted is overly aggressive and inefficient, as it essentially decreases the accuracy of $\mathcal{A}$.

Consequently, there is a need for more sophisticated methods to relax the conditions for accepting actions from $\mathcal{A}$. Potential solutions may include:
\begin{enumerate}
    \item Step-by-Step Relaxed Exact Match: While still performing step-by-step checks, this approach does not enforce an exact match between $\mathcal{A}$'s result and $\mathcal{T}$'s result. Instead, it allows for some degree of flexibility in what constitutes a match.
    \item Postponed Judgment: Instead of performing step-by-step checks, $\mathcal{T}$ will judge whether a sequence of $n$ steps proposed by $\mathcal{A}$ is within an acceptable range. This approach allows for a more holistic evaluation of $\mathcal{A}$'s proposals.
\end{enumerate}

By implementing these solutions, we can enhance the effectiveness and efficiency of speculative planning, making it more adaptable to the variability inherent in task completion.

\paragraph{Balancing time and cost efficiency in speculative planning}
The current version of speculative planning may incur a high additional cost. Therefore, balancing time efficiency and cost efficiency becomes a critical topic. Several methods can be employed to reduce the cost effectively:

\begin{enumerate}
    \item Utilize a Cost-Effective $\mathcal{A}$: Employ a cheaper yet well-functioning $\mathcal{A}$. This agent can be trained through knowledge distillation from $\mathcal{T}$, thereby improving performance while maintaining a smaller size.
    \item Enhance the Approximation-Target Judgment Method: Implement a more sophisticated approximation-target judgment method to improve the perceived accuracy of $\mathcal{A}$. This approach ensures that $\mathcal{A}$'s outputs are more reliably accepted, reducing the need for costly re-evaluations by $\mathcal{T}$.
\end{enumerate}

By implementing these methods, we can achieve a better balance between time efficiency and cost efficiency in speculative planning.

\paragraph{Limitations of the Current User Interface}
As mentioned in the algorithm design section, although the current user interface can handle active user input, these interactions and interruptions must be made ``on time.'' Specifically, users cannot change the result once it is fully presented in the user interface. This limitation means that users must closely monitor the algorithm's progress, and if they miss the opportunity to intervene, there is no way to go back and make modifications. In the future, the user interface should support backtracing to allow users to revisit and modify previous steps, enhancing the flexibility and usability of the system.

\section{Conclusion}
\label{sec:conclusion}
This paper introduces Interactive Speculative Planning, a novel approach that co-designs an efficient agent system with an active user interface to enhance the efficiency of agent planning while involving human interaction to further accelerate the system. By treating human interruptions as an integral part of the system, we not only make the planning process more user-centric but also accelerate the entire system by providing correct intermediate steps. Our experimental results on two benchmarks, OpenAGI and TravelPlanner, demonstrate the effectiveness of our approach in improving time efficiency and cost efficiency.

However, our work also highlights several limitations and future directions. The current system does not support backtracing, and the exact match method for accepting $\mathcal{A}$ actions is overly aggressive. Future work should focus on developing more sophisticated methods for relaxing the conditions for accepting actions from $\mathcal{A}$, enhancing security measures to mitigate the risks associated with speculative execution, and improving the user interface to support backtracing and provide greater flexibility for user interactions.

\section*{Appendix}
\subsection*{Implementation Details on TravelPlanner}
\label{tp}
To evaluate the final plans generated by normal agent planning and speculative planning, we adopt the metrics Delivery Rate and Commonsense Constraint Micro Pass Rate (the only two metrics with non-trivial results):
\paragraph{Delivery Rate} assesses whether agents can successfully deliver a final plan within a limited number of steps. Falling into dead loops, experiencing numerous failed attempts, or reaching the maximum number of steps (30 steps in our experimental setting) will result in failure.
\paragraph{Commonsense Constraint Micro Pass Rate} Commonsense Constraint Pass Rate comprises eight commonsense dimensions, which evaluates whether a language agent can incorporate commonsense into their plan without explicit instructions. The Macro Pass Rate indicates the ratio of passed constraints to the total number of constraints.

Below are the three set of results:

\paragraph{Setting 1} 
normal agent planning:\\
Delivery Rate: 55.6\%
Commonsense Constraint Micro Pass Rate: 48.6\%

speculative planning:\\
Delivery Rate: 55.6\%
Commonsense Constraint Micro Pass Rate: 41.7\%

\paragraph{Setting 2}

normal agent planning:\\
Delivery Rate: 55.6\%
Commonsense Constraint Micro Pass Rate: 41.7\%

speculative planning:\\
Delivery Rate: 55.6\%
Commonsense Constraint Micro Pass Rate: 34.7\%

\paragraph{Setting 3}

normal agent planning:\\
Delivery Rate: 55.6\%
Commonsense Constraint Micro Pass Rate: 54.3\%

speculative planning:\\
Delivery Rate: 55.6\%
Commonsense Constraint Micro Pass Rate: 48.6\%

\bibliography{references} 

\begin{thebibliography}{59}
\providecommand{\natexlab}[1]{#1}
\providecommand{\url}[1]{\texttt{#1}}
\expandafter\ifx\csname urlstyle\endcsname\relax
  \providecommand{\doi}[1]{doi: #1}\else
  \providecommand{\doi}{doi: \begingroup \urlstyle{rm}\Url}\fi

\bibitem[Barron et~al.(2004)Barron, Simpson, Rothrock, Frecker, Barton, and Ligetti]{barron2004graphical}
Kimberly Barron, Timothy~W Simpson, Ling Rothrock, Mary Frecker, Russell~R Barton, and Chris Ligetti.
\newblock Graphical user interfaces for engineering design: impact of response delay and training on user performance.
\newblock In \emph{International Design Engineering Technical Conferences and Computers and Information in Engineering Conference}, volume 46962, pp.\  11--20, 2004.

\bibitem[Cai et~al.(2024)Cai, Li, Geng, Peng, Lee, Chen, and Dao]{cai2024medusa}
Tianle Cai, Yuhong Li, Zhengyang Geng, Hongwu Peng, Jason~D Lee, Deming Chen, and Tri Dao.
\newblock Medusa: Simple llm inference acceleration framework with multiple decoding heads.
\newblock \emph{arXiv preprint arXiv:2401.10774}, 2024.

\bibitem[Carr et~al.(1992)Carr, Hasegawa, Lemmon, and Plaisant]{carr1992effects}
David Carr, Hiroaki Hasegawa, Doug Lemmon, and Catherine Plaisant.
\newblock The effects of time delays on a telepathology user interface.
\newblock In \emph{Proceedings of the Annual Symposium on Computer Application in Medical Care}, pp.\  256. American Medical Informatics Association, 1992.

\bibitem[Chan et~al.(2023)Chan, Chen, Su, Yu, Xue, Zhang, Fu, and Liu]{chan2023chateval}
Chi-Min Chan, Weize Chen, Yusheng Su, Jianxuan Yu, Wei Xue, Shanghang Zhang, Jie Fu, and Zhiyuan Liu.
\newblock Chateval: Towards better llm-based evaluators through multi-agent debate.
\newblock \emph{arXiv preprint arXiv:2308.07201}, 2023.

\bibitem[Chen et~al.(2023)Chen, Borgeaud, Irving, Lespiau, Sifre, and Jumper]{chen2023accelerating}
Charlie Chen, Sebastian Borgeaud, Geoffrey Irving, Jean-Baptiste Lespiau, Laurent Sifre, and John Jumper.
\newblock Accelerating large language model decoding with speculative sampling.
\newblock \emph{arXiv preprint arXiv:2302.01318}, 2023.

\bibitem[Chen et~al.(2024)Chen, Davis, Hanin, Bailis, Stoica, Zaharia, and Zou]{chen2024more}
Lingjiao Chen, Jared~Quincy Davis, Boris Hanin, Peter Bailis, Ion Stoica, Matei Zaharia, and James Zou.
\newblock Are more llm calls all you need? towards scaling laws of compound inference systems.
\newblock \emph{arXiv preprint arXiv:2403.02419}, 2024.

\bibitem[Cheng et~al.(2024)Cheng, Hu, Wang, Peng, Li, and Zhang]{cheng2024slice}
Ke~Cheng, Wen Hu, Zhi Wang, Hongen Peng, Jianguo Li, and Sheng Zhang.
\newblock Slice-level scheduling for high throughput and load balanced llm serving.
\newblock \emph{arXiv preprint arXiv:2406.13511}, 2024.

\bibitem[Ding et~al.(2024{\natexlab{a}})Ding, Mallick, Wang, Sim, Mukherjee, Ruhle, Lakshmanan, and Awadallah]{ding2024hybrid}
Dujian Ding, Ankur Mallick, Chi Wang, Robert Sim, Subhabrata Mukherjee, Victor Ruhle, Laks~VS Lakshmanan, and Ahmed~Hassan Awadallah.
\newblock Hybrid llm: Cost-efficient and quality-aware query routing.
\newblock \emph{arXiv preprint arXiv:2404.14618}, 2024{\natexlab{a}}.

\bibitem[Ding et~al.(2024{\natexlab{b}})Ding, Xu, and Lakshmanan]{ding2024occam}
Dujian Ding, Bicheng Xu, and Laks~VS Lakshmanan.
\newblock Occam: Towards cost-efficient and accuracy-aware image classification inference.
\newblock \emph{arXiv preprint arXiv:2406.04508}, 2024{\natexlab{b}}.

\bibitem[Du et~al.(2023)Du, Li, Torralba, Tenenbaum, and Mordatch]{du2023improving}
Yilun Du, Shuang Li, Antonio Torralba, Joshua~B Tenenbaum, and Igor Mordatch.
\newblock Improving factuality and reasoning in language models through multiagent debate.
\newblock \emph{arXiv preprint arXiv:2305.14325}, 2023.

\bibitem[Fan et~al.(2023)Fan, Hua, Li, Ling, Zhang, and Hemphill]{fan2023nphardeval}
Lizhou Fan, Wenyue Hua, Lingyao Li, Haoyang Ling, Yongfeng Zhang, and Libby Hemphill.
\newblock Nphardeval: Dynamic benchmark on reasoning ability of large language models via complexity classes.
\newblock \emph{arXiv preprint arXiv:2312.14890}, 2023.

\bibitem[Fan et~al.(2024)Fan, Hua, Li, Zhu, Jin, Li, Ling, Chi, Wang, Ma, et~al.]{fan2024nphardeval4v}
Lizhou Fan, Wenyue Hua, Xiang Li, Kaijie Zhu, Mingyu Jin, Lingyao Li, Haoyang Ling, Jinkui Chi, Jindong Wang, Xin Ma, et~al.
\newblock Nphardeval4v: A dynamic reasoning benchmark of multimodal large language models.
\newblock \emph{arXiv preprint arXiv:2403.01777}, 2024.

\bibitem[Gabbay \& Mendelson(1996)Gabbay and Mendelson]{gabbay1996speculative}
Freddy Gabbay and Avi Mendelson.
\newblock \emph{Speculative execution based on value prediction}.
\newblock Citeseer, 1996.

\bibitem[Ge et~al.(2023)Ge, Ren, Hua, Xu, Tan, and Zhang]{ge2023llm}
Yingqiang Ge, Yujie Ren, Wenyue Hua, Shuyuan Xu, Juntao Tan, and Yongfeng Zhang.
\newblock Llm as os, agents as apps: Envisioning aios, agents and the aios-agent ecosystem.
\newblock \emph{arXiv e-prints}, pp.\  arXiv--2312, 2023.

\bibitem[Ge et~al.(2024)Ge, Hua, Mei, Tan, Xu, Li, Zhang, et~al.]{ge2024openagi}
Yingqiang Ge, Wenyue Hua, Kai Mei, Juntao Tan, Shuyuan Xu, Zelong Li, Yongfeng Zhang, et~al.
\newblock Openagi: When llm meets domain experts.
\newblock \emph{Advances in Neural Information Processing Systems}, 36, 2024.

\bibitem[Hemmer et~al.(2023)Hemmer, Westphal, Schemmer, Vetter, V{\"o}ssing, and Satzger]{hemmer2023human}
Patrick Hemmer, Monika Westphal, Max Schemmer, Sebastian Vetter, Michael V{\"o}ssing, and Gerhard Satzger.
\newblock Human-ai collaboration: the effect of ai delegation on human task performance and task satisfaction.
\newblock In \emph{Proceedings of the 28th International Conference on Intelligent User Interfaces}, pp.\  453--463, 2023.

\bibitem[Hong et~al.(2023)Hong, Zheng, Chen, Cheng, Wang, Zhang, Wang, Yau, Lin, Zhou, et~al.]{hong2023metagpt}
Sirui Hong, Xiawu Zheng, Jonathan Chen, Yuheng Cheng, Jinlin Wang, Ceyao Zhang, Zili Wang, Steven Ka~Shing Yau, Zijuan Lin, Liyang Zhou, et~al.
\newblock Metagpt: Meta programming for multi-agent collaborative framework.
\newblock \emph{arXiv preprint arXiv:2308.00352}, 2023.

\bibitem[Horvitz(1999)]{horvitz1999principles}
Eric Horvitz.
\newblock Principles of mixed-initiative user interfaces.
\newblock In \emph{Proceedings of the SIGCHI conference on Human Factors in Computing Systems}, pp.\  159--166, 1999.

\bibitem[Hua et~al.(2023)Hua, Fan, Li, Mei, Ji, Ge, Hemphill, and Zhang]{hua2023war}
Wenyue Hua, Lizhou Fan, Lingyao Li, Kai Mei, Jianchao Ji, Yingqiang Ge, Libby Hemphill, and Yongfeng Zhang.
\newblock War and peace (waragent): Large language model-based multi-agent simulation of world wars.
\newblock \emph{arXiv preprint arXiv:2311.17227}, 2023.

\bibitem[Hua et~al.(2024)Hua, Yang, Li, Wei, and Zhang]{hua2024trustagent}
Wenyue Hua, Xianjun Yang, Zelong Li, Cheng Wei, and Yongfeng Zhang.
\newblock Trustagent: Towards safe and trustworthy llm-based agents through agent constitution.
\newblock \emph{arXiv preprint arXiv:2402.01586}, 2024.

\bibitem[Jawahar et~al.(2023)Jawahar, Abdul-Mageed, Lakshmanan, and Ding]{jawahar2023llm}
Ganesh Jawahar, Muhammad Abdul-Mageed, Laks~VS Lakshmanan, and Dujian Ding.
\newblock Llm performance predictors are good initializers for architecture search.
\newblock \emph{arXiv preprint arXiv:2310.16712}, 2023.

\bibitem[Jin et~al.(2024)Jin, Yu, Zhao, Hua, Meng, Zhang, Du, et~al.]{jin2024impact}
Mingyu Jin, Qinkai Yu, Haiyan Zhao, Wenyue Hua, Yanda Meng, Yongfeng Zhang, Mengnan Du, et~al.
\newblock The impact of reasoning step length on large language models.
\newblock \emph{arXiv preprint arXiv:2401.04925}, 2024.

\bibitem[Kocher et~al.(2020)Kocher, Horn, Fogh, Genkin, Gruss, Haas, Hamburg, Lipp, Mangard, Prescher, et~al.]{kocher2020spectre}
Paul Kocher, Jann Horn, Anders Fogh, Daniel Genkin, Daniel Gruss, Werner Haas, Mike Hamburg, Moritz Lipp, Stefan Mangard, Thomas Prescher, et~al.
\newblock Spectre attacks: Exploiting speculative execution.
\newblock \emph{Communications of the ACM}, 63\penalty0 (7):\penalty0 93--101, 2020.

\bibitem[Kwon et~al.(2024)Kwon, Hu, Myers, Karamcheti, Dragan, and Sadigh]{kwon2024toward}
Minae Kwon, Hengyuan Hu, Vivek Myers, Siddharth Karamcheti, Anca Dragan, and Dorsa Sadigh.
\newblock Toward grounded commonsense reasoning.
\newblock In \emph{2024 IEEE International Conference on Robotics and Automation (ICRA)}, pp.\  5463--5470. IEEE, 2024.

\bibitem[Leviathan et~al.(2023)Leviathan, Kalman, and Matias]{leviathan2023fast}
Yaniv Leviathan, Matan Kalman, and Yossi Matias.
\newblock Fast inference from transformers via speculative decoding.
\newblock In \emph{International Conference on Machine Learning}, pp.\  19274--19286. PMLR, 2023.

\bibitem[Li et~al.(2023)Li, Hammoud, Itani, Khizbullin, and Ghanem]{li2023camel}
Guohao Li, Hasan Hammoud, Hani Itani, Dmitrii Khizbullin, and Bernard Ghanem.
\newblock Camel: Communicative agents for" mind" exploration of large language model society.
\newblock \emph{Advances in Neural Information Processing Systems}, 36:\penalty0 51991--52008, 2023.

\bibitem[Li et~al.(2024)Li, Zhang, Yu, Fu, and Ye]{li2024more}
Junyou Li, Qin Zhang, Yangbin Yu, Qiang Fu, and Deheng Ye.
\newblock More agents is all you need.
\newblock \emph{arXiv preprint arXiv:2402.05120}, 2024.

\bibitem[Lin et~al.(2024)Lin, Hua, Li, Chang, Fan, Ji, Hua, Jin, Luo, and Zhang]{lin2024battleagent}
Shuhang Lin, Wenyue Hua, Lingyao Li, Che-Jui Chang, Lizhou Fan, Jianchao Ji, Hang Hua, Mingyu Jin, Jiebo Luo, and Yongfeng Zhang.
\newblock Battleagent: Multi-modal dynamic emulation on historical battles to complement historical analysis.
\newblock \emph{arXiv preprint arXiv:2404.15532}, 2024.

\bibitem[Liu et~al.(2024)Liu, Tang, Liu, Ni, Han, and Wang]{liu2024kangaroo}
Fangcheng Liu, Yehui Tang, Zhenhua Liu, Yunsheng Ni, Kai Han, and Yunhe Wang.
\newblock Kangaroo: Lossless self-speculative decoding via double early exiting.
\newblock \emph{arXiv preprint arXiv:2404.18911}, 2024.

\bibitem[Liu et~al.(2023{\natexlab{a}})Liu, Hu, Bailis, Stoica, Deng, Cheung, and Zhang]{liu2023online}
Xiaoxuan Liu, Lanxiang Hu, Peter Bailis, Ion Stoica, Zhijie Deng, Alvin Cheung, and Hao Zhang.
\newblock Online speculative decoding.
\newblock \emph{arXiv preprint arXiv:2310.07177}, 2023{\natexlab{a}}.

\bibitem[Liu et~al.(2023{\natexlab{b}})Liu, Zhang, Li, Liu, and Yang]{liu2023dynamic}
Zijun Liu, Yanzhe Zhang, Peng Li, Yang Liu, and Diyi Yang.
\newblock Dynamic llm-agent network: An llm-agent collaboration framework with agent team optimization.
\newblock \emph{arXiv preprint arXiv:2310.02170}, 2023{\natexlab{b}}.

\bibitem[Lubars \& Tan(2019)Lubars and Tan]{lubars2019ask}
Brian Lubars and Chenhao Tan.
\newblock Ask not what ai can do, but what ai should do: Towards a framework of task delegability.
\newblock \emph{Advances in neural information processing systems}, 32, 2019.

\bibitem[Mcilroy et~al.(2019)Mcilroy, Sevcik, Tebbi, Titzer, and Verwaest]{mcilroy2019spectre}
Ross Mcilroy, Jaroslav Sevcik, Tobias Tebbi, Ben~L Titzer, and Toon Verwaest.
\newblock Spectre is here to stay: An analysis of side-channels and speculative execution.
\newblock \emph{arXiv preprint arXiv:1902.05178}, 2019.

\bibitem[Mei et~al.(2024)Mei, Li, Xu, Ye, Ge, and Zhang]{mei2024llm}
Kai Mei, Zelong Li, Shuyuan Xu, Ruosong Ye, Yingqiang Ge, and Yongfeng Zhang.
\newblock Aios: Llm agent operating system.
\newblock \emph{arXiv preprint arXiv:2403.16971}, 2024.

\bibitem[Nakajima(2023)]{nakajima2023babyagi}
Yohei Nakajima.
\newblock Babyagi.
\newblock \emph{Retrieved April}, 25:\penalty0 2023, 2023.

\bibitem[Nightingale et~al.(2005)Nightingale, Chen, and Flinn]{nightingale2005speculative}
Edmund~B Nightingale, Peter~M Chen, and Jason Flinn.
\newblock Speculative execution in a distributed file system.
\newblock \emph{ACM SIGOPS operating systems review}, 39\penalty0 (5):\penalty0 191--205, 2005.

\bibitem[Oh et~al.(2024)Oh, Kim, Kim, Kim, Lee, Chang, and Seo]{oh2024exegpt}
Hyungjun Oh, Kihong Kim, Jaemin Kim, Sungkyun Kim, Junyeol Lee, Du-seong Chang, and Jiwon Seo.
\newblock Exegpt: Constraint-aware resource scheduling for llm inference.
\newblock In \emph{Proceedings of the 29th ACM International Conference on Architectural Support for Programming Languages and Operating Systems, Volume 2}, pp.\  369--384, 2024.

\bibitem[Qiao et~al.(2022)Qiao, Ou, Zhang, Chen, Yao, Deng, Tan, Huang, and Chen]{qiao2022reasoning}
Shuofei Qiao, Yixin Ou, Ningyu Zhang, Xiang Chen, Yunzhi Yao, Shumin Deng, Chuanqi Tan, Fei Huang, and Huajun Chen.
\newblock Reasoning with language model prompting: A survey.
\newblock \emph{arXiv preprint arXiv:2212.09597}, 2022.

\bibitem[Saha et~al.(2024)Saha, Prasad, Chen, Hase, Stengel-Eskin, and Bansal]{saha2024system}
Swarnadeep Saha, Archiki Prasad, Justin Chih-Yao Chen, Peter Hase, Elias Stengel-Eskin, and Mohit Bansal.
\newblock System-1. x: Learning to balance fast and slow planning with language models.
\newblock \emph{arXiv preprint arXiv:2407.14414}, 2024.

\bibitem[Shen et~al.(2024)Shen, Song, Tan, Li, Lu, and Zhuang]{shen2024hugginggpt}
Yongliang Shen, Kaitao Song, Xu~Tan, Dongsheng Li, Weiming Lu, and Yueting Zhuang.
\newblock Hugginggpt: Solving ai tasks with chatgpt and its friends in hugging face.
\newblock \emph{Advances in Neural Information Processing Systems}, 36, 2024.

\bibitem[Shinn et~al.(2024)Shinn, Cassano, Gopinath, Narasimhan, and Yao]{shinn2024reflexion}
Noah Shinn, Federico Cassano, Ashwin Gopinath, Karthik Narasimhan, and Shunyu Yao.
\newblock Reflexion: Language agents with verbal reinforcement learning.
\newblock \emph{Advances in Neural Information Processing Systems}, 36, 2024.

\bibitem[Simpson et~al.(2007)Simpson, Barron, Rothrock, Frecker, Barton, and Ligetti]{simpson2007impact}
Timothy~W Simpson, Kimberly Barron, Ling Rothrock, Mary Frecker, Russell~R Barton, and Chris Ligetti.
\newblock Impact of response delay and training on user performance with text-based and graphical user interfaces for engineering design.
\newblock \emph{Research in Engineering Design}, 18:\penalty0 49--65, 2007.

\bibitem[Spector \& Re(2023)Spector and Re]{spector2023accelerating}
Benjamin Spector and Chris Re.
\newblock Accelerating llm inference with staged speculative decoding.
\newblock \emph{arXiv preprint arXiv:2308.04623}, 2023.

\bibitem[Srivatsa et~al.(2024)Srivatsa, He, Abhyankar, Li, and Zhang]{srivatsa2024preble}
Vikranth Srivatsa, Zijian He, Reyna Abhyankar, Dongming Li, and Yiying Zhang.
\newblock Preble: Efficient distributed prompt scheduling for llm serving.
\newblock 2024.

\bibitem[Topsakal \& Akinci(2023)Topsakal and Akinci]{topsakal2023creating}
Oguzhan Topsakal and Tahir~Cetin Akinci.
\newblock Creating large language model applications utilizing langchain: A primer on developing llm apps fast.
\newblock In \emph{International Conference on Applied Engineering and Natural Sciences}, volume~1, pp.\  1050--1056, 2023.

\bibitem[Wang et~al.(2024)Wang, Wang, Su, Tong, and Song]{wang2024rethinking}
Qineng Wang, Zihao Wang, Ying Su, Hanghang Tong, and Yangqiu Song.
\newblock Rethinking the bounds of llm reasoning: Are multi-agent discussions the key?
\newblock \emph{arXiv preprint arXiv:2402.18272}, 2024.

\bibitem[Wei et~al.(2022)Wei, Wang, Schuurmans, Bosma, Xia, Chi, Le, Zhou, et~al.]{wei2022chain}
Jason Wei, Xuezhi Wang, Dale Schuurmans, Maarten Bosma, Fei Xia, Ed~Chi, Quoc~V Le, Denny Zhou, et~al.
\newblock Chain-of-thought prompting elicits reasoning in large language models.
\newblock \emph{Advances in neural information processing systems}, 35:\penalty0 24824--24837, 2022.

\bibitem[Wu et~al.(2023)Wu, Bansal, Zhang, Wu, Zhang, Zhu, Li, Jiang, Zhang, and Wang]{wu2023autogen}
Qingyun Wu, Gagan Bansal, Jieyu Zhang, Yiran Wu, Shaokun Zhang, Erkang Zhu, Beibin Li, Li~Jiang, Xiaoyun Zhang, and Chi Wang.
\newblock Autogen: Enabling next-gen llm applications via multi-agent conversation framework.
\newblock \emph{arXiv preprint arXiv:2308.08155}, 2023.

\bibitem[Wu et~al.(2024)Wu, Jia, Zhang, Li, Zhu, Wang, Lee, Peng, Wu, and Wang]{wu2024mathchat}
Yiran Wu, Feiran Jia, Shaokun Zhang, Hangyu Li, Erkang Zhu, Yue Wang, Yin~Tat Lee, Richard Peng, Qingyun Wu, and Chi Wang.
\newblock Mathchat: Converse to tackle challenging math problems with llm agents.
\newblock In \emph{ICLR 2024 Workshop on Large Language Model (LLM) Agents}, 2024.

\bibitem[Xi et~al.(2023)Xi, Chen, Guo, He, Ding, Hong, Zhang, Wang, Jin, Zhou, et~al.]{xi2023rise}
Zhiheng Xi, Wenxiang Chen, Xin Guo, Wei He, Yiwen Ding, Boyang Hong, Ming Zhang, Junzhe Wang, Senjie Jin, Enyu Zhou, et~al.
\newblock The rise and potential of large language model based agents: A survey.
\newblock \emph{arXiv preprint arXiv:2309.07864}, 2023.

\bibitem[Xie et~al.(2024)Xie, Zhang, Chen, Zhu, Lou, Tian, Xiao, and Su]{xie2024travelplanner}
Jian Xie, Kai Zhang, Jiangjie Chen, Tinghui Zhu, Renze Lou, Yuandong Tian, Yanghua Xiao, and Yu~Su.
\newblock Travelplanner: A benchmark for real-world planning with language agents.
\newblock \emph{arXiv preprint arXiv:2402.01622}, 2024.

\bibitem[Yao et~al.(2022)Yao, Zhao, Yu, Du, Shafran, Narasimhan, and Cao]{yao2022react}
Shunyu Yao, Jeffrey Zhao, Dian Yu, Nan Du, Izhak Shafran, Karthik Narasimhan, and Yuan Cao.
\newblock React: Synergizing reasoning and acting in language models.
\newblock \emph{arXiv preprint arXiv:2210.03629}, 2022.

\bibitem[Yao et~al.(2024)Yao, Yu, Zhao, Shafran, Griffiths, Cao, and Narasimhan]{yao2024tree}
Shunyu Yao, Dian Yu, Jeffrey Zhao, Izhak Shafran, Tom Griffiths, Yuan Cao, and Karthik Narasimhan.
\newblock Tree of thoughts: Deliberate problem solving with large language models.
\newblock \emph{Advances in Neural Information Processing Systems}, 36, 2024.

\bibitem[Zhang et~al.(2023{\natexlab{a}})Zhang, Krishna, Awadallah, and Wang]{zhang2023ecoassistant}
Jieyu Zhang, Ranjay Krishna, Ahmed~H Awadallah, and Chi Wang.
\newblock Ecoassistant: Using llm assistant more affordably and accurately.
\newblock \emph{arXiv preprint arXiv:2310.03046}, 2023{\natexlab{a}}.

\bibitem[Zhang et~al.(2023{\natexlab{b}})Zhang, Xia, Wang, Chen, Liu, Wu, and Liu]{zhang2023ideal}
Shaokun Zhang, Xiaobo Xia, Zhaoqing Wang, Ling-Hao Chen, Jiale Liu, Qingyun Wu, and Tongliang Liu.
\newblock Ideal: Influence-driven selective annotations empower in-context learners in large language models.
\newblock \emph{arXiv preprint arXiv:2310.10873}, 2023{\natexlab{b}}.

\bibitem[Zhang et~al.(2024{\natexlab{a}})Zhang, Zhang, Liu, Song, Wang, Krishna, and Wu]{zhang2024training}
Shaokun Zhang, Jieyu Zhang, Jiale Liu, Linxin Song, Chi Wang, Ranjay Krishna, and Qingyun Wu.
\newblock Training language model agents without modifying language models.
\newblock \emph{arXiv preprint arXiv:2402.11359}, 2024{\natexlab{a}}.

\bibitem[Zhang et~al.(2024{\natexlab{b}})Zhang, Zheng, Li, Yang, Li, Wang, Chao, Wang, Li, and Ji]{zhang2024you}
Shaokun Zhang, Xiawu Zheng, Guilin Li, Chenyi Yang, Yuchao Li, Yan Wang, Fei Chao, Mengdi Wang, Shen Li, and Rongrong Ji.
\newblock You only compress once: Towards effective and elastic bert compression via exploit--explore stochastic nature gradient.
\newblock \emph{Neurocomputing}, 599:\penalty0 128140, 2024{\natexlab{b}}.

\bibitem[Zhang et~al.(2024{\natexlab{c}})Zhang, Mao, Ge, Wang, de~Wynter, Xia, Wu, Song, Lan, and Wei]{zhang2024llm}
Yadong Zhang, Shaoguang Mao, Tao Ge, Xun Wang, Adrian de~Wynter, Yan Xia, Wenshan Wu, Ting Song, Man Lan, and Furu Wei.
\newblock Llm as a mastermind: A survey of strategic reasoning with large language models.
\newblock \emph{arXiv preprint arXiv:2404.01230}, 2024{\natexlab{c}}.

\bibitem[Zhou et~al.(2024)Zhou, Ning, Hong, Fu, Xu, Li, Lou, Wang, Yuan, Li, et~al.]{zhou2024survey}
Zixuan Zhou, Xuefei Ning, Ke~Hong, Tianyu Fu, Jiaming Xu, Shiyao Li, Yuming Lou, Luning Wang, Zhihang Yuan, Xiuhong Li, et~al.
\newblock A survey on efficient inference for large language models.
\newblock \emph{arXiv preprint arXiv:2404.14294}, 2024.

\end{thebibliography}
\bibliographystyle{iclr2024_conference}

\end{document}